\documentclass{jfm}
\usepackage{graphicx}
\usepackage{epstopdf, epsfig,amsmath,amssymb}

\shorttitle{Tidal turbines in an infinitely large array}

\shortauthor{P. Ouro and T. Nishino}
\title{Performance and wake characteristics of tidal turbines in an infinitely large array}

\author{Pablo Ouro\aff{1,2} \corresp{\email{pablo.ouro@manchester.ac.uk}} \and Takafumi Nishino\aff{3}}
\affiliation{
\aff{1}Department of Mechanical, Aerospace and Civil Engineering, The University of Manchester, Manchester M13 9PL, UK
\aff{2}Hydro-environmental Research Centre, School of Engineering, Cardiff University, The Parade, Cardiff CF24 3AA, UK
\aff{3}Department of Engineering Science, University of Oxford, Parks Road, Oxford OX1 3PJ, UK
}

\begin{document}

\maketitle

\begin{abstract}
The efficiency of tidal-stream turbines in a large array depends on the balance between negative effects of turbine-wake interactions and positive effects of bypass-flow acceleration due to local blockage, both of which are functions of the layout of turbines. In this study we investigate the hydrodynamics of turbines in an infinitely large array with aligned or staggered layouts for a range of streamwise and lateral turbine spacing. First, we present a theoretical analysis based on an extension of the Linear Momentum Actuator Disc Theory (LMADT) for perfectly aligned and staggered layouts, employing a hybrid inviscid-viscous approach to account for the local blockage effect within each row of turbines and the viscous (turbulent) wake mixing behind each row in a coupled manner. We then present Large-Eddy Simulation (LES) of open-channel flow for 28 different layouts of tidal turbines using an Actuator-Line Method (ALM) with doubly periodic boundary conditions. Both theoretical and LES results show that the efficiency of turbines (or the power of turbines for a given bulk velocity) in an aligned array decreases as we reduce the streamwise turbine spacing, whereas that in a staggered array remains high and may even increase due to the positive local blockage effect (causing the local flow velocity upstream of each turbine to exceed the bulk velocity) if the lateral turbine spacing is sufficiently small. The LES results further reveal that the amplitude of wake meandering tends to decrease as we reduce the lateral turbine spacing, which leads to a lower wake recovery rate in the near-wake region. These results will help to understand and improve the efficiency of tidal turbines in future large tidal arrays, even though the performance of real tidal arrays may depend not only on turbine-to-turbine interactions within the array but also on macro-scale interactions between the array and natural tidal currents, the latter of which are outside the scope of this study.

\end{abstract}

\begin{keywords}
%coastal engineering; wakes %tidal turbines, wake recovery, array layout,
\end{keywords}

\section{Introduction}

Tidal stream energy is being developed rapidly as a promising complement to intermittent wind and solar energy, owing to the high predictability of tides \citep{Adcock2021}. To ensure tidal energy is successfully embedded into the future net-zero carbon energy mix, we need to understand the flow physics driving the efficiency of tidal turbine arrays when deployed at a large scale. Due to the confined flow environment in which these turbines operate, we cannot simply infer tidal array hydrodynamics from existing knowledge of wind farm aerodynamics \citep{Porte-Agel2020,Nishino2020}. For instance, relatively shallow waters may restrict the ability of turbine wakes to expand vertically, potentially affecting the recovery of wakes in a large turbine array.

Future tidal arrays will comprise multiple rows of turbines, whose design requires to consider an appropriate spacing between turbines as well as their impact on the tidal channel flow dynamics.
These micro- (turbine to turbine) and macro-scale (array to tidal channel) interactions need to be considered when designing tidal arrays in order to find the optimal trade-off between energy extraction and minimal changes to the tidal flow \citep{DeDominicis2018}.
Large tidal arrays deployed at a tidal channel lead to an added resistance to its flow dynamics, which, if too large, can considerably obstruct and modify the overall flow. Therefore, to optimise the design of large tidal arrays, it is required to tune the operating conditions of individual turbines for a given tidal channel (e.g. straight or variable-section), tidal forcing and turbine density \citep{Vennell2010,Vennell2011,Vennell2015}.

At the micro-scale, the power production capability of tidal turbines is driven by turbulent wake mixing and acceleration of bypass flow in-between turbines. In relatively closely packed arrays, limiting the negative wake-turbine interactions is often key to minimising power losses \citep{Stallard2013}. However, local blockage due to a small lateral spacing between devices (as well as a shallow water depth confining the flow) may lead to local flow acceleration that enhances individual turbine power, as observed in experimental tests, e.g. \citet{Stallard2013,Noble2020}. The effect of local blockage has also been investigated using the Linear Momentum Actuator Disc Theory (LMADT) for a single lateral row of turbines \citep{Garrett2007,Nishino2012,Nishino2013,Vogel2016,Creed2017} and two rows of turbines \citep{Draper2014}. These studies suggest that two staggered rows of turbines tend to be more efficient than two perfectly aligned (or centred) rows of turbines, but less efficient than a single row of the same total number of turbines with the same array width.

Whilst the above findings are important for the performance of arrays with a small number of rows, future tidal arrays will require turbines to be deployed in several rows to generate a sufficiently large amount of energy, which may cause macro-scale flow interactions between the array and the tidal channel. \citet{Vennell2010,Vennell2011} combined the LMADT with a simple theoretical tidal channel flow model to analyse how the resistance and lateral spacing of turbines within each row should be tuned for a given number of rows deployed across a given tidal channel, to maximise the total power generation. However, existing theoretical tidal array models based on the LMADT, including the two-row model of \citet{Draper2014}, do not fully account for the complex effect of turbulent wake mixing. The Vennell-type array models assume that the streamwise spacing between rows is large enough for individual turbine wakes to be fully mixed after each row, whereas the two-row model of \citet{Draper2014} assumes that the two rows are close enough to each other for the effect of wake mixing to be negligible.

In narrow tidal channels or straits, turbines in a multi-row array may operate in a fully-waked scenario (similar to perfectly aligned layout) during half of the tidal cycle (e.g. ebb tide), but they may operate in partly-waked conditions (akin to staggered layouts) during the other half of the cycle (e.g. flood tide) \citep{GarciaNovo2018}. This asymmetry between ebb and flood tides further complicates the design of optimal array configurations, requiring to understand the performance and hydrodynamics of a given array design for various incident flow characteristics. Many existing studies looking into tidal array optimisation have adopted low-fidelity flow models, such as two-dimensional shallow water models \citep{Culley2016}, analytical wake models, e.g. Gaussian models \citep{Stansby2015} and aforementioned theoretical models based on the LMADT \citep{Nishino2013,Draper2014}, while high-fidelity simulations have been restricted to relatively small arrays with a limited number of configurations, due to their large computational expense \citep{Afgan2013,Chawdhary2017,Ouro2019JFS}.
However, as the performance of tidal devices in arrays is driven by wake-turbine interactions as well as bathymetry-induced turbulence \citep{Stallard2013,Ouro2018FTC}, turbulence-resolving approaches such as Large-Eddy Simulation (LES) are valuable to yield reliable hydrodynamics results as well as to build more accurate low-order models that can improve array optimisation tools.

In this paper, we investigate the flow characteristics and efficiency of infinitely large tidal arrays with perfectly aligned and staggered configurations, combining predictions from two contrasting approaches: actuator disc theory and LES. Infinitely large arrays represent an asymptotic case in which the flow passing the turbine rows is fully developed (i.e. flow statistics become identical for all rows). This is similar to large wind farms in which such flow conditions may be attained approximately after 10 to 15 rows, depending on atmospheric stability conditions \citep{Porte-Agel2020,Sharma2018}. We consider a wide range of streamwise and lateral turbine spacing to understand how the array efficiency can be maximised (from the micro-scale perspective) by balancing the negative impact of turbine wakes impinging downstream turbines and the positive local blockage effects. The paper is structured as follows: in \S\ref{sec:theory} we introduce an extended theoretical model developed for periodic turbine arrays with perfectly aligned and staggered configurations. Details of 28 LES runs comprising aligned and staggered layouts are presented in \S\ref{sec:testcases}, with results of flow characteristics and hydrodynamic coefficients in \S\ref{sec:results}. In \S\ref{sec:discussion} we provide further discussion on the comparison between the predictions obtained from the theoretical analysis and LES, followed by main conclusions in \S\ref{sec:conclusion}.

%%%%%%%%%%%%%%%%%%%%%%%%%%
%\newpage \pagebreak \vspace{5cm}

\section{Theoretical analysis}\label{sec:theory}
We start with a simple theoretical analysis on the efficiency of an infinitely large array of ideal turbines in a steady, uniform and vertically confined flow. The analysis is based on the work of \citet{Draper2014}, who extended the LMADT for laterally confined flows \citep{Garrett2007,Houlsby2008} to predict an upper limit to the efficiency of two aligned or staggered rows of turbines. This two-row analysis is further extended in this study to investigate an infinite number of aligned or staggered rows of tidal turbines, following the idea of hybrid inviscid-viscous approach recently proposed by \citet{Nishino2019}.

\begin{figure}
\vspace{0.2cm}
\centerline{\includegraphics[width=.9\linewidth]{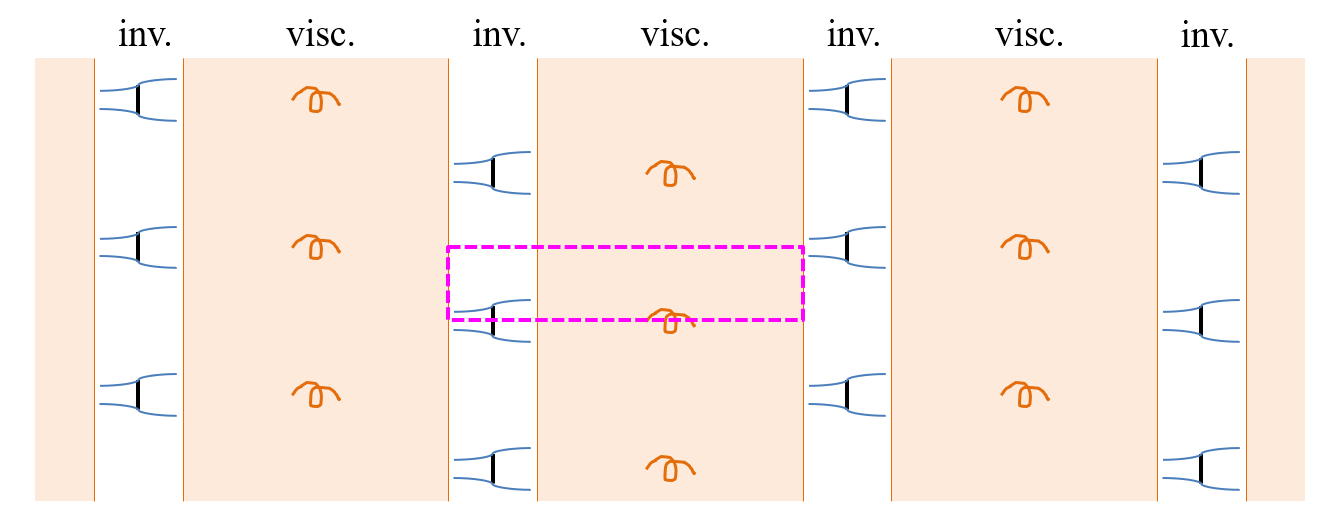}}
\caption{Schematic of the flow past a periodic staggered array of tidal turbines, divided into the hypothetical inviscid (inv.) and viscous (visc.) flow zones. The rectangular region enclosed by the magenta dashed line corresponds to that depicted in figure \ref{fig:theory2}.} \label{fig:theory1}
\end{figure}

Figure \ref{fig:theory1} illustrates an example of a periodic staggered array of tidal turbines. The key assumption employed in the hybrid inviscid-viscous approach is that the streamwise extent of the region in which the expansion of the flow through each turbine takes place is much shorter than the distance between each row of turbines. With this assumption, we hypothetically divide the flow field into two types of zones; namely the inviscid flow zones, which are analysed using the LMADT neglecting the effect of viscous (or turbulent) mixing, and the viscous flow zones, which are modelled separately to account for the effect of mixing. This approach is also in line with the results of a recent LES study of flow past a periodic array of actuator discs \citep{West2020} showing that the effects of inviscid and viscous (turbulent) flow processes are dominant, respectively, in the vicinity of the turbines and in the rest of the flow field.

\begin{figure}
\vspace{0.2cm}
\centerline{\includegraphics[width=.9\linewidth]{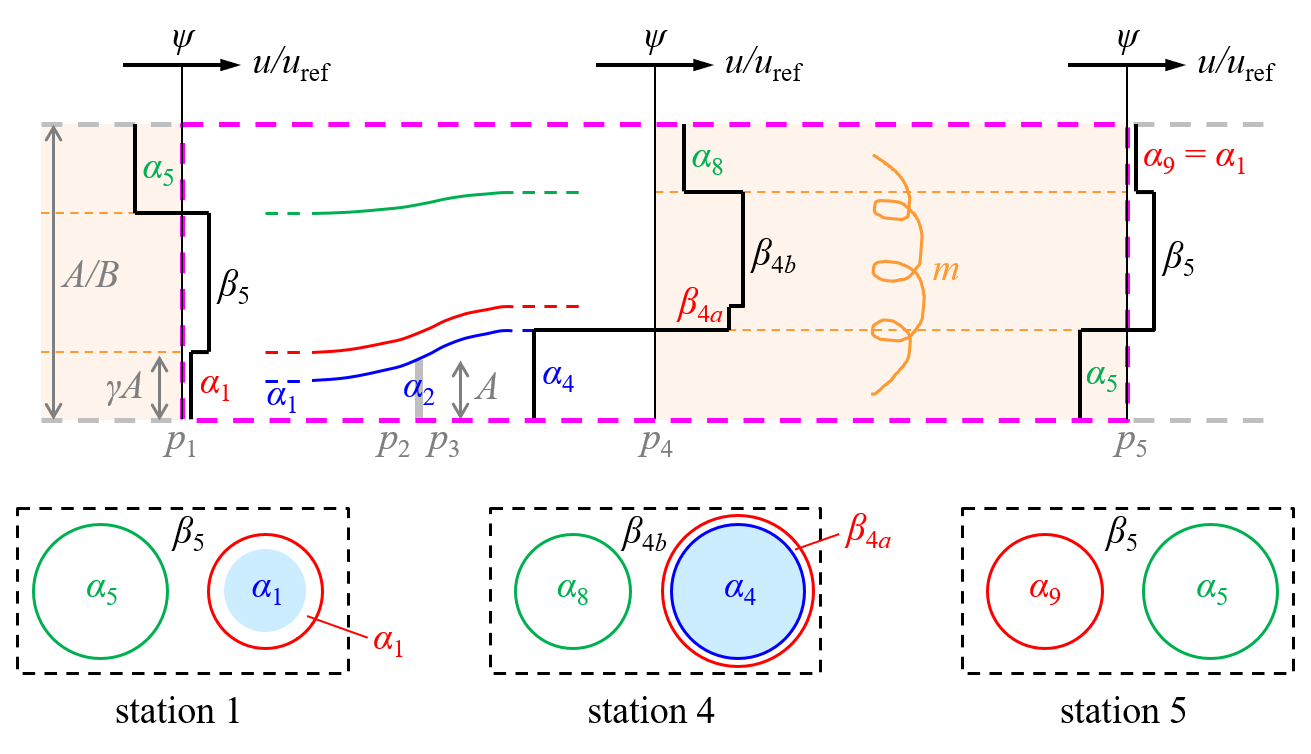}}
\caption{Schematic of the quasi-1D theoretical model for a periodic staggered array of tidal turbines, with examples of how cross-sectional flow patterns may appear in a corresponding 3D flow problem. The three vertical thin black lines indicate the locations of stations 1, 4 and 5 as well as the (non-dimensional) cross-sectional average velocity, $\psi = u_\mathrm{av}/u_\mathrm{ref}$, for the superposed plot of $u/u_\mathrm{ref}$ at each station, whereas the thick black lines show the profiles of $u/u_\mathrm{ref}$ at the three stations (calculated for the case with $B=0.2$, $K=3$ and $m=0.7$). Reproduced from \citet{Nishino2019} with minor modifications.} \label{fig:theory2}
\end{figure}

\subsection{Staggered rows of actuator discs}\label{sec:theory_s}
A schematic of the theoretical model for the staggered case is shown in figure \ref{fig:theory2}. Here we consider a straight local flow passage containing only one-half of an actuator disc due to the periodic and symmetric nature of the array as shown in figure \ref{fig:theory1}. The cross-sectional area of the flow passage is $A/B$, where $A$ is the half-disc area and $B$ is the area blockage ratio. Although the figure is depicted in a two-dimensional manner, this is still a quasi-one-dimensional flow model as we do not consider any variation of flow quantities (velocity and pressure) over the cross-section of each streamtube in the inviscid zone. Assuming that the flow is incompressible, the average velocity over the cross-section of the entire flow passage, $u_\mathrm{av}$, does not change in the streamwise direction.

Following the common notations used in the LMADT \citep{Houlsby2008} we define four stations within the inviscid zone: station 1 is at the inlet of the inviscid zone where the core-flow streamtube (that encompasses the actuator disc) starts to expand, stations 2 and 3 are immediately upstream and downstream of the disc, respectively, and station 4 is at the outlet of the inviscid zone where the pressure is equalised between the core- and bypass-flow streamtubes. In addition, we describe the outlet of the viscous zone as station 5 (which is station 1 for the next row of discs).  The pressure at stations 1 to 5 is denoted by $p_1$ to $p_5$, respectively, whereas the velocity is described using the velocity coefficients $\alpha$ and $\beta$ with subscripts as in figure \ref{fig:theory2}. Each velocity coefficient represents the ratio of the velocity there to a reference velocity, $u_\mathrm{ref}$. In the following we take the core-flow velocity at station 1 as $u_\mathrm{ref}$ (i.e. $\alpha_1 = 1$) for convenience.

We now consider the conservation of mass, momentum and energy in the inviscid flow zone in a similar manner to the work of \citet{Draper2014}. First, since the Bernoulli equation must be satisfied to conserve energy in each of the three bypass-flow streamtubes between stations 1 and 4, we obtain
\begin{eqnarray}
&& p_1-p_4 = \frac{1}{2}\rho u_\mathrm{ref}^2 \left( \beta_{4a}^2-1 \right), \label{eq:te1}\\
&& p_1-p_4 = \frac{1}{2}\rho u_\mathrm{ref}^2 \left( \beta_{4b}^2-\beta_5^2 \right), \label{eq:te2}\\
&& p_1-p_4 = \frac{1}{2}\rho u_\mathrm{ref}^2 \left( \alpha_8^2-\alpha_5^2 \right), \label{eq:te3}
\end{eqnarray}
where $\rho$ is the fluid density. Substituting (\ref{eq:te2}) and (\ref{eq:te3}), respectively, into (\ref{eq:te1}) gives
\begin{eqnarray}
&& \beta_{4b} = \left( \beta_{4a}^2 + \beta_{5}^2 - 1 \right)^{1/2}, \label{eq:tb4b}\\
&& \alpha_{8} = \left( \beta_{4a}^2 + \alpha_{5}^2 - 1 \right)^{1/2}. \label{eq:ta8}
\end{eqnarray}
As the Bernoulli equation must also be satisfied in the upstream part (between stations 1 and 2) and downstream part (between stations 3 and 4) of the core-flow streamtube, we also obtain
\begin{eqnarray}
p_2-p_3 = p_1-p_4 + \frac{1}{2}\rho u_\mathrm{ref}^2 \left( 1-\alpha_4^2 \right), \label{eq:te4}
\end{eqnarray}
which, together with (\ref{eq:te1}), leads to an expression for the half-disc thrust, $T$, as
\begin{eqnarray}
T = (p_2-p_3)A = \frac{1}{2}\rho u_\mathrm{ref}^2 A \left( \beta_{4a}^2 - \alpha_4^2 \right), \label{eq:tthrust}
\end{eqnarray}
and thus an expression for the half-disc power, $P$, as
\begin{eqnarray}
P = T\alpha_2 u_\mathrm{ref} = \frac{1}{2}\rho u_\mathrm{ref}^3 A \alpha_2\left( \beta_{4a}^2 - \alpha_4^2 \right). \label{eq:tpower}
\end{eqnarray}
Hence, to obtain $P$ for a given $\alpha_2$, for example, we need to know how $\alpha_4$ and $\beta_{4a}$ depend on $\alpha_2$. By considering the conservation of mass in the `main' bypass-flow streamtube between station 1 (where the velocity is $\beta_5 u_\mathrm{ref}$) and station 4 (where the velocity is $\beta_{4b} u_\mathrm{ref}$) we obtain
\begin{eqnarray}
\beta_5 u_\mathrm{ref} A \left(\frac{1}{B} - \frac{\alpha_2}{\alpha_4} - \gamma \right) = \beta_{4b} u_\mathrm{ref} A \left(\frac{1}{B} - \frac{\alpha_2}{\alpha_4} - \frac{\gamma - \alpha_2}{\beta_{4a}} - \gamma \right), \label{eq:tmass}
\end{eqnarray}
which leads to
\begin{eqnarray}
\alpha_4 = \frac{\alpha_2}{\dfrac{1}{B} - \dfrac{\beta_{4b}}{\beta_{4a}} \left( \dfrac{\gamma - \alpha_2}{\beta_{4b} - \beta_5} \right)  - \gamma}, \label{eq:ta4}
\end{eqnarray}
where
\begin{eqnarray}
\gamma = \frac{\alpha_2 \alpha_5}{\alpha_4 \alpha_8} \label{eq:tgamma}
\end{eqnarray}
is the area ratio for the wake flow upstream of the disc as depicted in figure \ref{fig:theory2}. Note that this wake originates from the disc located two rows upstream (and its velocity increases from $\alpha_5 u_\mathrm{ref}$ to $\alpha_8 u_\mathrm{ref}$ when it passes through the row immediately upstream) due to the staggered configuration. As for $\beta_{4a}$ we consider the conservation of momentum for the entire flow passage to obtain
\begin{equation}
\begin{split}
& (p_1 - p_4)\frac{A}{B} - T \\
& = \rho u_\mathrm{ref}^2 A \left[ \alpha_2  (\alpha_4 - 1) + (\gamma - \alpha_2)(\beta_{4a} - 1) + \gamma \alpha_8 (\alpha_8 - \alpha_5) + \beta_5 \left( \frac{1}{B} - \frac{\alpha_2}{\alpha_4} -\gamma \right) \left( \beta_{4b} - \beta_5 \right) \right] , \label{eq:tmom}
\end{split}
\end{equation}
which, together with (\ref{eq:te1}) and (\ref{eq:tthrust}), leads to
\begin{equation}
\begin{split}
\left( 1-B \right)\beta_{4a}^2 - 2B \left( \gamma-\alpha_2 \right) \beta_{4a} &- 2B \left[ \alpha_2 \alpha_4 - \frac{1}{2}\alpha_4^2 + \gamma \left( \alpha_8^2 - \alpha_5 \alpha_8 - 1 \right) \right] \\
&-2\beta_5 \left( \beta_{4b} - \beta_5 \right) \left[ 1-B \left( \frac{\alpha_2}{\alpha_4} + \gamma \right) \right] - 1 = 0. \label{eq:tb4a}
\end{split}
\end{equation}

The above set of equations for the inviscid zone is not closed as it includes $\alpha_5$ and $\beta_5$, which need to be modelled considering the effect of mixing in the viscous zone. There are several possible ways to model the effect of mixing but here we employ a very simple approach using a single non-dimensional input parameter called the mixing factor, $m$, which represents the completeness of mixing in the viscous zone. Specifically, the value of $m$ (between 0 and 1) describes how much the velocity of flow at each cross-sectional position returns to the cross-sectional average velocity of the entire flow passage, $u_\mathrm{av}$, as the flow passes through the viscous zone. By applying this to the wake flow of the disc we obtain
\begin{eqnarray}
\alpha_5 = m\psi + \left( 1-m \right) \alpha_4 , \label{eq:ta5}
\end{eqnarray}
where $\psi = u_\mathrm{av}/u_\mathrm{ref}$ is the non-dimensional cross-sectional average velocity and this can be calculated from the velocity profile at station 1, for example, as
\begin{eqnarray}
\psi = B \left[ \gamma \left( \alpha_8 + 1 \right) + \beta_5 \left( \frac{1}{B} - \frac{\alpha_2}{\alpha_4} - \gamma \right) \right] . \label{eq:tpsi}
\end{eqnarray}
For the bypass flow, however, the difficulty is that the number of flow passages at station 4 does not agree with that at station 1, since the actuator disc creates the additional narrow bypass streamtube that has $\beta_{4a}$ at station 4. To obtain a closed set of equations for this periodic flow problem within the framework of quasi-one-dimensional modelling, here we assume that the narrow bypass flow immediately outside of the wake is `merged' or fully mixed with the main bypass flow regardless of the value of $m$. This assumption seems reasonable since the difference between $\beta_{4a}$ and $\beta_{4b}$ tends to be small as depicted in figure \ref{fig:theory2}, and eventually leads to
\begin{eqnarray}
\beta_5 = m\psi + \left( 1-m \right) \beta_{4m} , \label{eq:tb5}
\end{eqnarray}
where $\beta_{4m}$ is the `area-weighted' average of $\beta_{4a}$ and $\beta_{4b}$, i.e.,
\begin{eqnarray}
\beta_{4m} = \frac{\left( \gamma - \alpha_2 \right) + \beta_5 \left( \dfrac{1}{B} - \dfrac{\alpha_2}{\alpha_4} - \gamma \right)}{\dfrac{\left( \gamma - \alpha_2 \right)}{\beta_{4a}} + \dfrac{\beta_5}{\beta_{4b}} \left( \dfrac{1}{B} - \dfrac{\alpha_2}{\alpha_4} - \gamma \right)} . \label{eq:tb4m}
\end{eqnarray}
It should be noted that $\alpha_9 = m\psi + \left( 1-m \right) \alpha_8$ is required together with (\ref{eq:ta5}) to (\ref{eq:tb4m}) to conserve the total mass in the viscous zone, but this is automatically satisfied as we enforce $\alpha_9=\alpha_1=1$ in this analysis due to the periodicity of the flow.

Finally, the thrust and power coefficients of the disc are expressed (for a given cross-sectional average velocity of the entire flow, $u_\mathrm{av}=\psi u_\mathrm{ref}$) as
\begin{eqnarray}
&& C_T = \frac{T}{\frac{1}{2}\rho \left( \psi u_\mathrm{ref} \right)^2 A} = \frac{\beta_{4a}^2 - \alpha_4^2}{\psi^2} , \label{eq:tct}\\
&& C_P = \frac{P}{\frac{1}{2}\rho \left( \psi u_\mathrm{ref} \right)^3 A} = \frac{\alpha_2\left( \beta_{4a}^2 - \alpha_4^2 \right)}{\psi^3} . \label{eq:tcp}
\end{eqnarray}
For convenience, we also define the resistance coefficient (or local thrust coefficient) of the disc as
\begin{eqnarray}
K = \frac{T}{\frac{1}{2}\rho \left( \alpha_2 u_\mathrm{ref} \right)^2 A} , \label{eq:tk}
\end{eqnarray}
from which and (\ref{eq:tthrust}), we obtain
\begin{eqnarray}
\alpha_2 = \left( \frac{\beta_{4a}^2 - \alpha_4^2}{K} \right)^{1/2} . \label{eq:ta2}
\end{eqnarray}

In summary, the above theoretical model for an infinite number of staggered rows of identical ideal turbines consists of three non-dimensional input parameters ($B$, $K$, $m$), ten non-dimensional unknowns to be determined ($\alpha_2$, $\alpha_4$, $\alpha_5$, $\alpha_8$, $\beta_{4a}$, $\beta_{4b}$, $\beta_{4m}$, $\beta_5$, $\gamma$, $\psi$) and a set of ten equations to be solved numerically: (\ref{eq:tb4b}), (\ref{eq:ta8}), (\ref{eq:ta4}), (\ref{eq:tgamma}), (\ref{eq:tb4a}), (\ref{eq:ta5}), (\ref{eq:tpsi}), (\ref{eq:tb5}), (\ref{eq:tb4m}) and (\ref{eq:ta2}).

\subsection{Aligned rows of actuator discs}\label{sub:theory_a}
Compared to the staggered case, the theoretical model for the aligned case becomes simpler as only two bypass-flow streamtubes (instead of three) need to be considered in the inviscid zone. This is because the wake encounters the disc immediately downstream (instead of two rows downstream). In the following, we again consider the conservation of mass, momentum and energy in the inviscid zone, and the simplified mixing process in the viscous zone, to derive a similar but smaller set of equations for the aligned case.

By applying the Bernoulli equation to the two bypass-flow streamtubes and the core-flow streamtube in the same manner as in the staggered case, we obtain (\ref{eq:te1}), (\ref{eq:te2}) and (\ref{eq:te4}), which lead to (\ref{eq:tb4b}), (\ref{eq:tthrust}) and (\ref{eq:tpower}) for $\beta_{4b}$, $T$ and $P$, respectively. Note that these equations are identical for both aligned and staggered cases, whereas (\ref{eq:te3}) and (\ref{eq:ta8}) are only for the staggered case (since the third bypass-flow streamtube does not exist in the aligned case). Next, from the conservation of mass in the `main' bypass-flow streamtube, we obtain
\begin{eqnarray}
\beta_5 u_\mathrm{ref} A \left(\frac{1}{B} - \frac{\alpha_2}{\alpha_4} \right) = \beta_{4b} u_\mathrm{ref} A \left[\frac{1}{B} - \frac{\alpha_2}{\alpha_4} - \frac{1}{\beta_{4a}}\left( \frac{\alpha_2}{\alpha_4}-\alpha_2 \right) \right], \label{eq:t2mass}
\end{eqnarray}
which leads to an expression for $\alpha_4$ as
\begin{eqnarray}
\alpha_4 = \frac{\alpha_2 \left( 1 + \dfrac{1}{\beta_{4a}} - \dfrac{\beta_5}{\beta_{4b}} \right)}{\dfrac{1}{B}\left( 1 - \dfrac{\beta_5}{\beta_{4b}} \right) + \dfrac{\alpha_2}{\beta_{4a}} }. \label{eq:t2a4}
\end{eqnarray}
For $\beta_{4a}$ we consider the conservation of momentum for the entire flow to obtain
\begin{equation}
\begin{split}
& (p_1 - p_4)\frac{A}{B} - T \\
& = \rho u_\mathrm{ref}^2 A \left[ \alpha_2  (\alpha_4 - 1) + \left( \frac{\alpha_2}{\alpha_4} - \alpha_2 \right)(\beta_{4a} - 1) + \beta_5 \left( \frac{1}{B} - \frac{\alpha_2}{\alpha_4} \right) \left( \beta_{4b} - \beta_5 \right) \right] , \label{eq:t2mom}
\end{split}
\end{equation}
which, together with (\ref{eq:te1}) and (\ref{eq:tthrust}), leads to
\begin{equation}
\begin{split}
\left( 1-B \right)\beta_{4a}^2 - 2B \left( \frac{\alpha_2}{\alpha_4}-\alpha_2 \right) \beta_{4a} &- 2B \left( \alpha_2 \alpha_4 - \frac{1}{2}\alpha_4^2 - \frac{\alpha_2}{\alpha_4} \right) \\
&-2\beta_5 \left( \beta_{4b} - \beta_5 \right) \left( 1-B \frac{\alpha_2}{\alpha_4} \right) - 1 = 0. \label{eq:t2b4a}
\end{split}
\end{equation}
For $\beta_5$ we need to consider the effect of mixing in the viscous zone. By employing the same approach as in the staggered case, we obtain (\ref{eq:tb5}) for the aligned case as well. Note, however, that $\psi$ and $\beta_{4m}$ are different for the aligned case, which are
\begin{eqnarray}
\psi = B \left[ \frac{\alpha_2}{\alpha_4} + \beta_5 \left( \frac{1}{B} - \frac{\alpha_2}{\alpha_4} \right) \right] , \label{eq:t2psi}
\end{eqnarray}
and
\begin{eqnarray}
\beta_{4m} = \frac{\left( \dfrac{\alpha_2}{\alpha_4} - \alpha_2 \right) + \beta_5 \left( \dfrac{1}{B} - \dfrac{\alpha_2}{\alpha_4} \right)}{\dfrac{1}{\beta_{4a}}\left( \dfrac{\alpha_2}{\alpha_4} - \alpha_2 \right) + \dfrac{\beta_5}{\beta_{4b}} \left( \dfrac{1}{B} - \dfrac{\alpha_2}{\alpha_4} \right)} . \label{eq:t2b4m}
\end{eqnarray}
We also need (\ref{eq:ta5}) together with the above equations to conserve the total mass in the viscous zone, but this is automatically satisfied since here we enforce $\alpha_5=\alpha_1=1$ due to the periodicity of the flow. Finally, $C_T$, $C_P$ and $K$ are all defined as in the staggered case, yielding the same equations (\ref{eq:tct}) to (\ref{eq:tk}) and thus (\ref{eq:ta2}) for $\alpha_2$.

In summary, the theoretical model for the aligned case consists of  three non-dimensional input parameters ($B$, $K$, $m$), seven non-dimensional unknowns ($\alpha_2$, $\alpha_4$, $\beta_{4a}$, $\beta_{4b}$, $\beta_{4m}$, $\beta_5$, $\psi$) and a set of seven equations to be solved numerically: (\ref{eq:tb4b}), (\ref{eq:tb5}), (\ref{eq:ta2}), (\ref{eq:t2a4}), (\ref{eq:t2b4a}), (\ref{eq:t2psi}) and (\ref{eq:t2b4m}).

\subsection{Example solutions}\label{sub:theory_e}
Here we present some example solutions of the above theoretical model, first for the aligned case and then for the staggered case. As this is a three parameter ($B$-$K$-$m$) problem, we start with fixing the disc resistance coefficient, $K$, at its optimal value for the complete mixing case, i.e., $m=1$. When $m=1$, our problem (for both aligned and staggered cases) reduces to the two-parameter ($B$-$K$) problem of \citet{Garrett2007} (hereafter referred to as GC07); hence, this optimal value for $m=1$ is known to be $K=2(1+B)^3/(1-B)^2$, which we refer to as $K_\mathrm{GC07}$.

\begin{figure}
\vspace{0.2cm}
\centerline{\includegraphics[width=.7\linewidth]{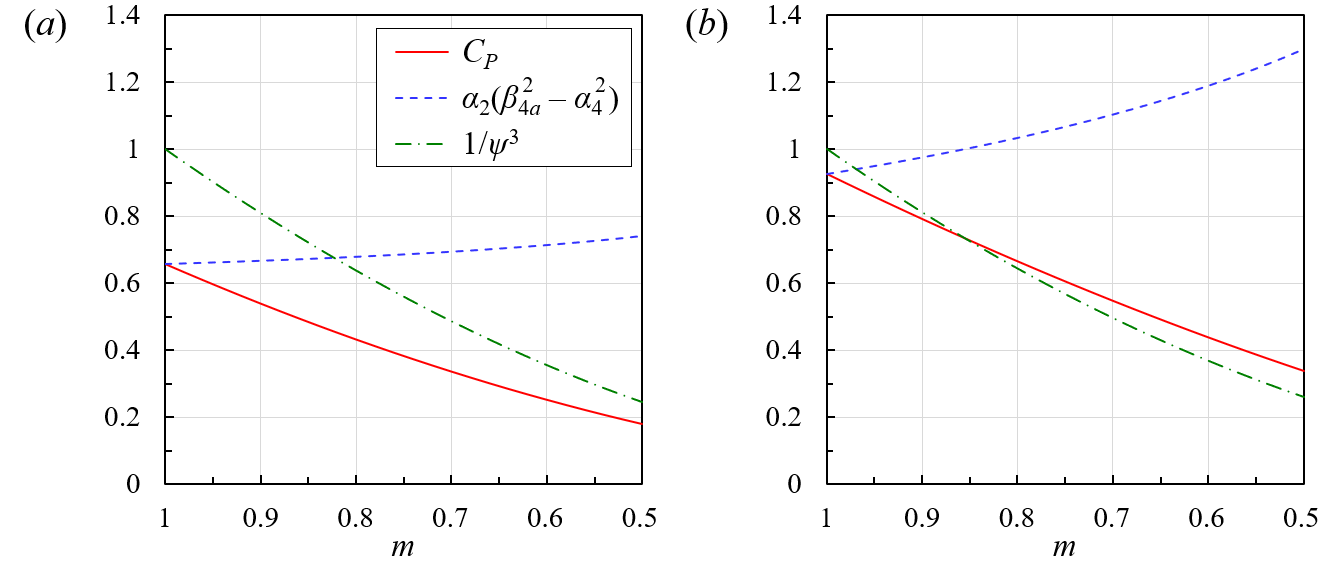}}
\caption{Theoretical prediction of the effect of mixing on the performance of aligned rows of actuator discs at two different blockage ratios: ($a$) $B=0.05$ and ($b$) $B=0.2$. The disc resistance coefficient is fixed at the optimal value for the case with complete mixing ($m=1$), i.e., $K=2(1+B)^3/(1-B)^2$.} \label{fig:theory3}
\end{figure}

Figure \ref{fig:theory3} shows how the performance of aligned rows of such `suboptimal' actuator discs (with $K=K_\mathrm{GC07}$) changes with $m$, at two different blockage ratios, $B=0.05$ and $0.2$. As can be expected intuitively, the power coefficient $C_P$ of aligned discs decreases monotonically as the mixing factor $m$ decreases (regardless of the blockage ratio). This agrees with the common observation that the power of aligned rows of turbines tends to decrease as we reduce the streamwise spacing between the rows, noting that the mixing tends to be less complete (i.e., $m$ tends to be small) when the spacing is small.

We also present in figure \ref{fig:theory3} the values of $\alpha_2(\beta_{4a}^2-\alpha_4^2)$ and $1/\psi^3$, to explain why $C_P$ decreases as $m$ decreases. As can be seen from (\ref{eq:tcp}), $C_P$ is equivalent to the product of these two values; the former is a different power coefficient defined using the velocity upstream of the disc ($u_\mathrm{ref}$) instead of the cross-sectional average velocity ($u_\mathrm{av}$), and the latter represents the effect of the difference between $u_\mathrm{ref}$ and $u_\mathrm{av}$. Here we can see that the value of $\alpha_2(\beta_{4a}^2-\alpha_4^2)$ actually increases as $m$ decreases. This is essentially because the local blockage effect is enhanced (or the `effective' blockage ratio increases) when the upstream core flow is slower than the upstream bypass flow \citep{Draper2016}. This enhancement of the local blockage effect caused by the incomplete wake mixing of the upstream disc, however, is not strong enough to compensate for the `loss' of power possessed by the upstream core flow compared to the `original' power possessed by the cross-sectionally averaged flow (i.e., the increase rate of $\alpha_2(\beta_{4a}^2-\alpha_4^2)$ is smaller than the decrease rate of $1/\psi^3$); therefore, $C_P$ eventually decreases as $m$ decreases.

\begin{figure}
\vspace{0.2cm}
\centerline{\includegraphics[width=1.\linewidth]{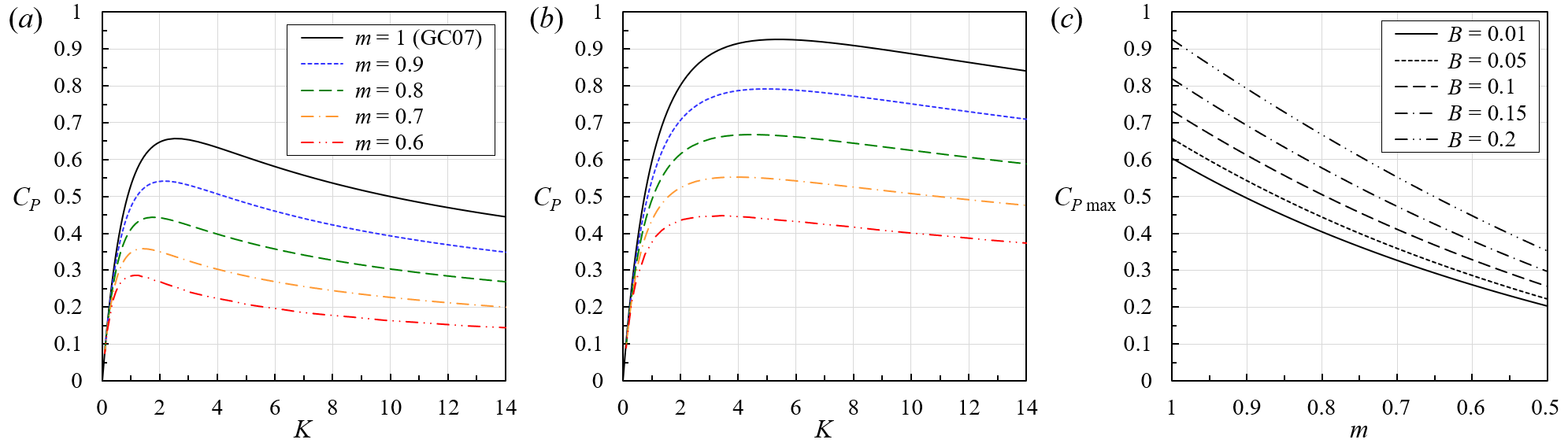}}
\caption{Theoretical $C_P$ versus $K$ for the aligned case at ($a$) $B=0.05$ and ($b$) $B=0.2$; and ($c$) the maximum $C_P$ obtained by varying $K$ for a given set of $B$ and $m$.} \label{fig:theory4}
\end{figure}

Although the above analysis was for `suboptimal' discs with $K=K_\mathrm{GC07}$, the same relationship between $C_P$ and $m$ is generally found for aligned discs with any given $K$ values. Figures \ref{fig:theory4}($a$) and ($b$) show $C_P$ versus $K$ curves for five different $m$ values, again at $B=0.05$ and $0.2$, respectively, demonstrating the trend. It can also be seen that the optimal $K$ value (to maximise $C_P$) tends to decrease as $m$ decreases, although the $C_P$ obtained at $K=K_\mathrm{GC07}$ is still close to the maximum $C_P$ for a given $m$ (especially when the blockage ratio is high). Figure \ref{fig:theory4}($c$) shows how the maximum $C_P$, or $C_{P\mathrm{max}}$, changes with $m$. As can be expected, for aligned rows of actuator discs, $C_{P\mathrm{max}}$ also decreases monotonically as $m$ decreases.

\begin{figure}
\vspace{0.2cm}
\centerline{\includegraphics[width=1.\linewidth]{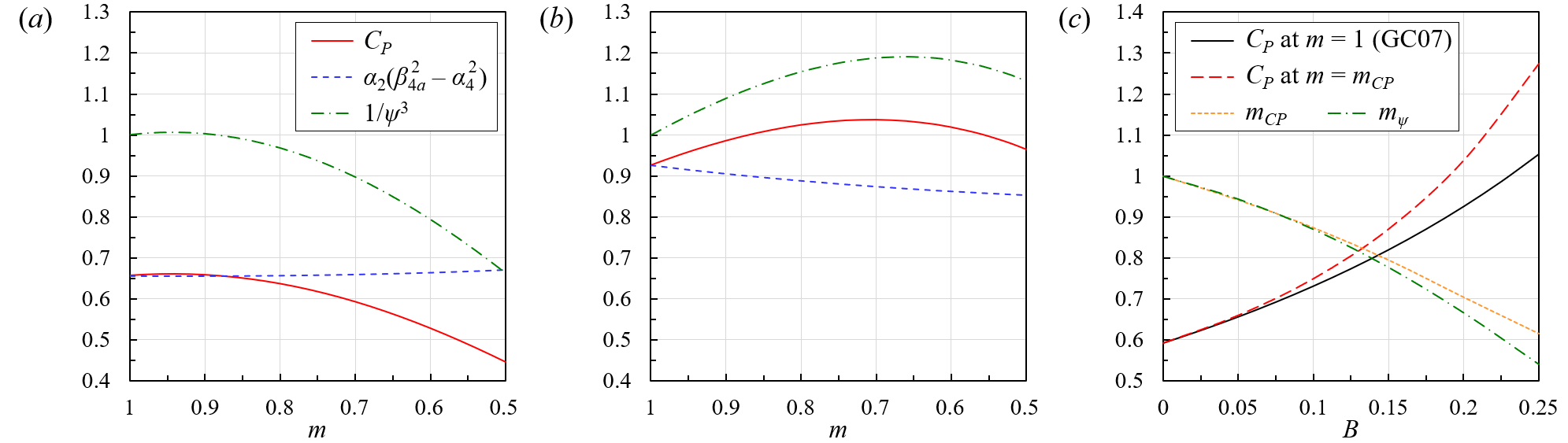}}
\caption{Theoretical prediction of the performance of staggered rows of actuator discs: the effect of mixing at ($a$) $B=0.05$ and ($b$) $B=0.2$; and ($c$) comparison of $C_P$ versus $B$ between the complete mixing case ($m=1$) and optimal mixing case ($m=m_{C_P}$). Plotted together are $m_{C_P}$ and $m_{\psi}$, which are the values of $m$ to maximise $C_P$ and to minimise $\psi$, respectively. The disc resistance coefficient is fixed at the optimal value for $m=1$, i.e., $K=2(1+B)^3/(1-B)^2$.} \label{fig:theory5}
\end{figure}

Next, we focus on staggered rows of actuator discs. Similarly to the aligned case, we start with the performance of `suboptimal' discs with $K=K_\mathrm{GC07}$, which is shown in figures \ref{fig:theory5}($a$) and ($b$) for $B=0.05$ and $B=0.2$, respectively. At $B=0.05$, the value of $1/\psi^3$ slightly increases (meaning that the upstream core flow velocity becomes slightly higher than the cross-sectional average velocity) as $m$ decreases from 1 to about 0.95, and then decreases as $m$ further decreases. In contrast, the value of $\alpha_2(\beta_{4a}^2-\alpha_4^2)$ first decreases slightly and then increases as $m$ decreases from 1; this is again due to changes in the `effective' blockage ratio, i.e., the local blockage effect is enhanced (or diminished) when the upstream core flow is surrounded by a faster (or slower) bypass flow. However, the increase (or decrease) rate of $\alpha_2(\beta_{4a}^2-\alpha_4^2)$ tends to be smaller than the decrease (or increase) rate of $1/\psi^3$, and hence $C_P$ follows the trend of $1/\psi^3$, i.e., $C_P$ slightly increases first and then decreases as $m$ decreases from 1. This initial increase in $1/\psi^3$ (and thus in $C_P$) is more clearly seen at $B=0.2$, demonstrating that the beneficial effect of the staggered layout (allowing the velocity upstream of the disc to become higher than the cross-sectional average velocity) is enhanced at a higher blockage ratio.

Figure \ref{fig:theory5}($c$) compares $C_P$ versus $B$ curves for the complete mixing case ($m=1$) and for the `optimal mixing' case ($m=m_{C_P}$, where $m_{C_P}$ is the value of $m$ that maximises $C_P$), again for staggered rows of `suboptimal' discs ($K=K_\mathrm{GC07}$). Also plotted together are $m_{C_P}$ and $m_{\psi}$, the latter of which represents the value of $m$ that minimises $\psi$ (and thus maximises $1/\psi^3$). It can be seen how the increase in $C_P$ due to the beneficial effect of the staggered layout is enhanced, and how the optimal level of mixing decreases, as we increase the blockage ratio. It is also worth noting that the difference between $m_{C_P}$ and $m_{\psi}$ is small, especially at low blockage ratios. This reflects the dominant influence of $1/\psi^3$ on $C_P$ as described earlier in figures \ref{fig:theory5}($a$) and ($b$).

\begin{figure}
\vspace{0.2cm}
\centerline{\includegraphics[width=1.\linewidth]{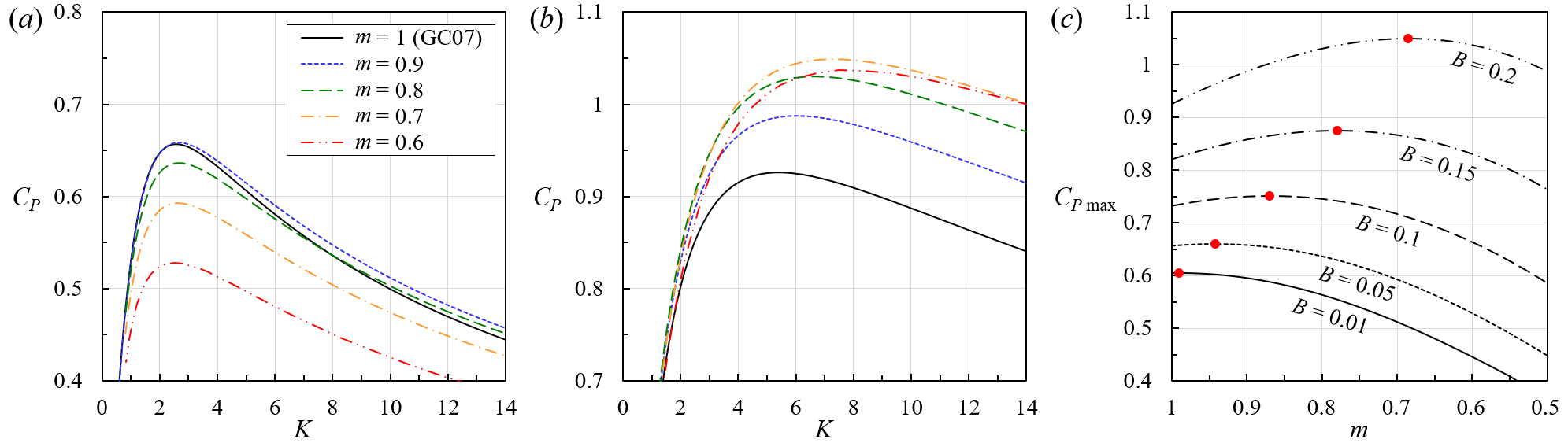}}
\caption{Theoretical $C_P$ versus $K$ for the staggered case at ($a$) $B=0.05$ and ($b$) $B=0.2$; and ($c$) the maximum $C_P$ obtained by varying $K$ for a given set of $B$ and $m$.} \label{fig:theory6}
\end{figure}

The $C_P$ values presented above are for `suboptimal' discs (with $K=K_\mathrm{GC07}$) and can therefore be increased further by adjusting $K$ for a given $m$ (or for a given streamwise distance between the rows). However, this additional increase in $C_P$ achieved by varying $K$ is rather small for the staggered case. Figures \ref{fig:theory6}($a$) and ($b$) show $C_P$ versus $K$ curves for five different $m$ values, again at $B=0.05$ and $0.2$, respectively. When $B$ is small, $K_\mathrm{GC07}$ tends to be already close to the optimal $K$ value for a given $m$; hence, the benefit of further adjusting $K$ is small. When $B$ is large, $K_\mathrm{GC07}$ is not so close to the optimal $K$ for a given $m$, but the $C_P$ versus $K$ curve tends to have a flatter peak; hence again, the additional gain in $C_P$ by adjusting $K$ is small. Figure \ref{fig:theory6}($c$) shows the maximum $C_P$ (achieved by adjusting $K$) versus $m$ curves for five different blockage ratios. The curves for $B=0.05$ and $0.2$ are in fact very similar to the $C_P$ versus $m$ curves for $K=K_\mathrm{GC07}$ plotted earlier in figures \ref{fig:theory5}($a$) and ($b$). At $B=0.2$, for example, the highest $C_P$ value achieved with $K=K_\mathrm{GC07}$ is only 1.1$\%$ lower than that achieved by optimising $K$.


In summary, these example solutions of the simple three-parameter ($B$-$K$-$m$) theoretical model suggest the following: 
(i) For an infinitely large staggered rows of tidal turbines, there is an optimal streamwise turbine spacing (for a given cross-sectional blockage ratio or a given lateral turbine spacing) to maximise the power of each turbine relative to the power of cross-sectional average flow. This is an outcome of the combined effects of local blockage and wake mixing, i.e., this optimum exists because the local flow velocity upstream of each disc can become higher than the cross-sectional average velocity (and this does help to increase the relative power despite reducing the `effective' blockage ratio) when the streamwise turbine spacing is reasonably (not excessively) large for the wake mixing behind each disc to be largely (but not entirely) completed. 
(ii) The optimal turbine resistance to maximise this relative power also depends on both lateral and streamwise turbine spacing, but the optimal value for a single row, namely $K=2(1+B)^3/(1-B)^2$, is expected to yield close to the maximum relative power. 
(iii) For an infinitely large aligned rows of turbines, however, this relative power is maximised simply when the streamwise turbine spacing is large enough for the wake mixing to be entirely completed; hence, the optimal turbine resistance becomes identical to that for a single row.

Finally, it should be remembered that, in the real world, the array of turbines cannot be infinitely large and that the cross-sectional average velocity (which was considered as a fixed parameter in the above analysis) will depend on how the flow resistance caused by the entire array alters the natural tidal-channel-scale momentum balance (see, e.g., \citet{Vennell2012, Vennell2015,Gupta2017}). Therefore, the true optimal turbine resistance (to maximise the power of turbines in a given tidal channel) will differ from that discussed in the above analysis.

%%%%%%%%%%%%%
%\newpage \pagebreak \vspace{5cm}
\section{Large-eddy simulation set-up}\label{sec:testcases}

The theoretical model presented above has been developed on the assumption that the flow around each turbine is inviscid and steady. However, real turbine wakes are unsteady with coherent flow structures developed at a rotor level, e.g. tip-vortices, and further downstream, e.g. wake meandering. Thus, we now investigate how the theoretical predictions of the performance of infinitely large tidal arrays compare to high-fidelity numerical predictions using Large-Eddy Simulation (LES). We adopt the well-validated in-house LES code Digital Offshore Farms Simulator (DOFAS) in which turbine blades are represented using an Actuator Line Method (ALM) \citep{Ouro2019JFS} and the flow solver is fully parallelised using the Message Passing Interface (MPI), providing a great computational scalability and performance \citep{Ouro2019CAF}. Details of the flow solver are provided in the Appendix. % \ref{app:DOFAS}.

We intentionally set the flow conditions to be as similar as possible to the conditions considered in the theoretical analysis, whilst ensuring that the physical dimensions of the turbines are close to those found in real tidal stream turbines. The 1MW DEEPGen IV tidal stream turbine design from the ReDAPT project is adopted, and the details of the blade hydrodynamic data used for the ALM are available in \cite{Scarlett2019}. The turbines have a diameter ($D$) of 12 m, rotating at a constant speed that corresponds to a tip-speed ratio of 4.0 (which is known to be optimal for the case of single-turbine operation), and include 10 m long ($0.8D$) nacelles. For convenience, we set the cross-sectional average velocity $U_0$ to 2.0 m s$^{-1}$ and consider this as our reference velocity, which yields a rotational speed of $\Omega = 1.35$ rad s$^{-1}$ and a full-revolution period of $T$ = 4.724 s.
%The corresponding Reynolds number based on the turbine diameter is 2.4$\cdot 10^6$.
%This LES computational setup intends to be as similar as possible to the conditions considered in the LMADT theoretical model whilst the physical dimensions are close to real-size tidal stream turbine farms. 

The computational domain is 432 m long ($L$), 144 m wide ($W$) and 24 m high ($H$), equivalent to $36D \times 12D \times 2D$, which is close to $6\pi H \times 2\pi H \times H$ commonly used in turbulent channel-flow simulations. Turbines are vertically centred at mid-water depth, i.e. $z=H/2$, to reduce vertical asymmetry effects that may complicate the comparison between LMADT and LES. 
A fixed time step $\Delta t$ is set to 0.045 s together with a uniform spatial resolution $\Delta x$ equal to 0.25 m, yielding a total of 48 mesh elements across the rotor diameter which is similar to the resolution adopted in other LES-ALM studies, e.g. \citet{Churchfield2013,Yang2019,Foti2019}.
Hence, the domain is divided into 1692 \(\times\) 576 \(\times\) 96 grid cells over the three spatial directions with a total of about 93.5 $\times$ 10$^6$ elements. 
Simulations are run using 864 processors on three High-Performance Computing facilities, namely Supercomputing Wales, GW4 Isambard, and ARCHER. 

 %, thus 105 time steps are required to compute a single revolution. 

\begin{figure}
\centerline{\includegraphics[width=.95\linewidth]{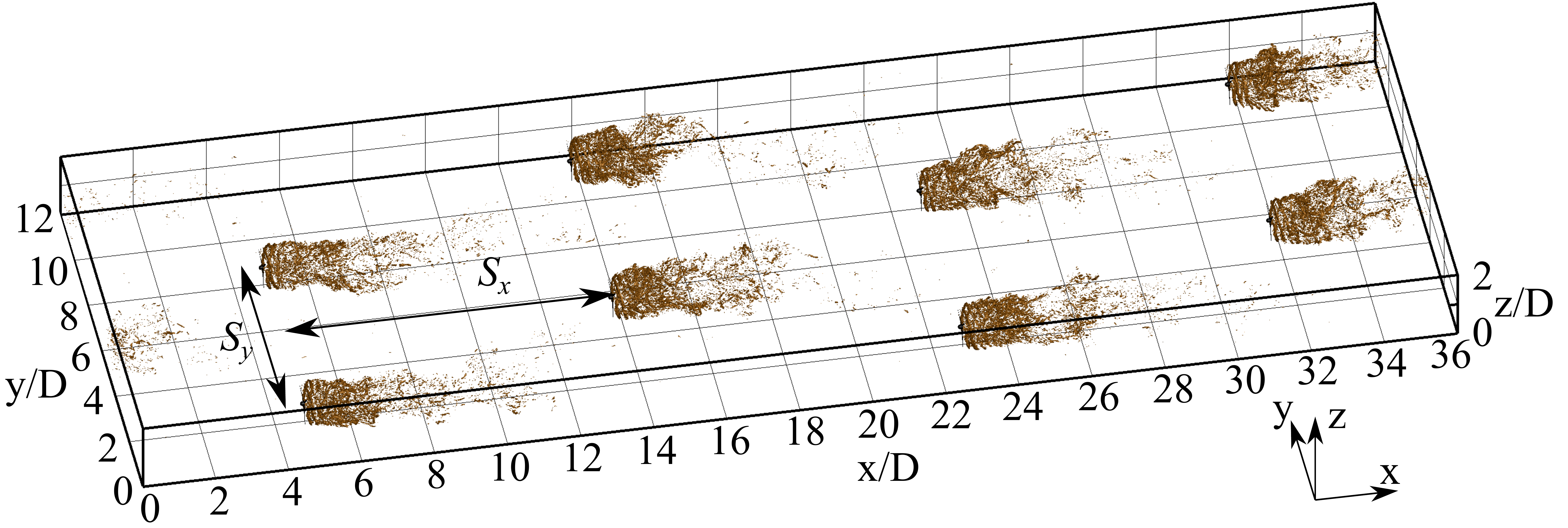}}
\caption{Dimensions of the computational domain and instantaneous flow structures generated in the ST-9x6 case, visualised using iso-surfaces of the Q-criterion.} \label{fig:compdomain}
\end{figure}

In the present LES, we represent infinite turbine arrays by imposing periodic boundary conditions in the streamwise and transverse directions. The flow is driven by the pressure gradient term $\Pi_i$ which enforces the mass flux across the entire domain to be constant, providing the cross-sectional averaged velocity is equal to $U_0$. Fixing the bulk velocity enables a direct comparison between the LES and the theoretical analysis in \S\ref{sec:theory}, unlike fixing $\Pi_i$ as in other infinitely-large wind farm simulations \citep{Calaf2010,Sharma2018}. Note that, to simulate a real large tidal array, both of the bulk velocity and $\Pi_i$ would need to vary depending on the macro-scale momentum balance over the entire tidal channel, which is outside the scope of the present study.
Shear-free conditions are adopted at the top boundary to represent a free surface, whilst the bottom shear stress is calculated using wall functions for a hydrodynamically smooth wall, similarly to \citet{Kang2014}. A representation of the computational domain is presented in figure \ref{fig:compdomain} together with instantaneous flow structures visualised using Q-criterion iso-surfaces for one of the configurations studied.

An initial precursor simulation was performed for over 60 eddy turnover time units ($t_e$), corresponding to nearly 27 flow-through ($FT = L / U_0$), in order to generate fully-developed turbulent flow conditions to be used as initial flow field for each array simulation. We perform 28 infinitely-large array simulations, whose details are presented in table \ref{table:setup} providing values of normalised streamwise and transverse separation between turbines ($S_x/D$, $S_y/D$), local blockage ratio $B$ (relating the turbine's projected area to the open-channel cross-section and number of turbines per row, i.e. $B = n_y \pi D^2 /4 H W$), physical simulated time in terms of flow-through time, and the number of computed revolutions. Also presented in this table are the key results of each simulation, namely the thrust coefficient ($C_T$), power coefficient ($C_P$), their relative difference ($\Delta C_T$, $\Delta C_P$) from the AL-36x12 configuration as a reference case, and the resistance coefficient ($K$). %friction velocity, ($u_*= \sqrt{d p/d x H}$, with $d p/d x$ being the streamwise pressure gradient)
The arrays adopted in the LES are perfectly aligned and staggered layouts labelled as {AL} and {ST} followed by the device spacing $S_x$x$S_y$, e.g. case {ST-9x6} corresponds to a staggered array with $S_x/D$ = 9 and $S_y/D$ = 6 as in figure \ref{fig:compdomain}.
%For completeness, we performed an additional simulation comprising a single turbine (labelled as "isolated" turbine case) with inflow-outflow boundary conditions with a 1/7$^{th}$ power law inlet velocity distribution \citep{Ouro2017JFS}, to assess the impact of streamwise periodic flow conditions in the wake and turbine performance.
%Flow statistics are computed after three and seven $FT$ for first- and second-order statistics, respectively, being in most cases second-order statistics averaged for at least 20$FT$, deemed sufficient to obtain converged statistics. 

\begin{table}
  \begin{center}
\def~{\hphantom{0}}
\begin{tabular}{lcccccccccccc} 
Case & $S_x/D$ & $S_y/D$ & $B$ & Time ($FT$)  & Revs  & $C_T$ & $C_P$ & $\Delta C_T$   & $\Delta C_P$ & $K$ \\
%Precursor  & -    & -    & -  & -  & -     & 5800     & 26.9      & 1246 & -     & -     & -     & -      \\
%Isolated & -    & -    & 1  & 1  & 0.033 & 5107     & 23.6      & 1097 & 0.773 & 0.298 & 1.008 & 1.041  \\
%\\
AL-36x12 & 36   & 12   & 0.033 & 27.1      & 1257 & 0.767 & 0.287 & 1.000 & 1.000 & 2.55  \\
AL-18x12 & 18   & 12   & 0.033 & 27.1      & 1257 & 0.766 & 0.277 & 0.999 & 0.965 & 2.74  \\
AL-9x12  & 9    & 12   & 0.033 & 27.1      & 1257 & 0.751 & 0.264 & 0.979 & 0.922 & 2.80  \\
AL-6x12  & 6    & 12   & 0.033 & 20.6      & 957  & 0.737 & 0.252 & 0.961 & 0.879 & 2.82  \\
ST-18x12 & 18   & 12   & 0.033 & 25.0      & 1160 & 0.777 & 0.292 & 1.014 & 1.021 & 2.29  \\
ST-9x12  & 9    & 12   & 0.033 & 25.0      & 1160 & 0.774 & 0.291 & 1.009 & 1.016 & 2.18  \\
ST-6x12  & 6    & 12   & 0.033 & 25.0      & 1160 & 0.774 & 0.292 & 1.009 & 1.019 & 2.07  \\
\\
AL-36x6  & 36   & 6    & 0.065 & 24.6      & 1141 & 0.786 & 0.301 & 1.025 & 1.050 & 2.20  \\
AL-18x6  & 18   & 6    & 0.065 & 26.6      & 1233 & 0.785 & 0.302 & 1.024 & 1.054 & 2.11  \\
AL-9x6   & 9    & 6    & 0.065 & 20.9      & 972  & 0.772 & 0.293 & 1.006 & 1.021 & 2.12  \\
AL-6x6   & 6    & 6    & 0.065 & 20.8      & 967  & 0.764 & 0.285 & 0.996 & 0.996 & 2.02  \\
ST-18x6  & 18   & 6    & 0.065 & 24.8      & 1151 & 0.784 & 0.300 & 1.022 & 1.045 & 2.28  \\
ST-9x6   & 9    & 6    & 0.065 & 23.4      & 1088 & 0.783 & 0.299 & 1.020 & 1.045 & 2.25  \\
ST-6x6   & 6    & 6    & 0.065 & 25.0      & 1160 & 0.777 & 0.293 & 1.013 & 1.024 & 2.27  \\
\\
AL-36x4  & 36   & 4    & 0.098 & 24.7      & 1146 & 0.787 & 0.307 & 1.026 & 1.073 & 1.92  \\
AL-18x4  & 18   & 4    & 0.098 & 25.0      & 1160 & 0.787 & 0.310 & 1.026 & 1.082 & 1.89  \\
AL-9x4   & 9    & 4    & 0.098 & 23.3      & 1083 & 0.777 & 0.300 & 1.013 & 1.046 & 1.88  \\
AL-6x4   & 6    & 4    & 0.098 & 23.4      & 1088 & 0.765 & 0.285 & 0.997 & 0.995 & 2.08  \\
ST-18x4  & 18   & 4    & 0.098 & 23.1      & 1074 & 0.787 & 0.310 & 1.027 & 1.082 & 1.93  \\
ST-9x4   & 9    & 4    & 0.098 & 23.4      & 1088 & 0.781 & 0.314 & 1.018 & 1.094 & 1.82  \\
ST-6x4   & 6    & 4    & 0.098 & 19.9      & 924  & 0.780 & 0.312 & 1.017 & 1.089 & 1.78  \\
\\
AL-36x3  & 36   & 3    & 0.131 & 27.1      & 1257 & 0.796 & 0.316 & 1.038 & 1.102 & 2.09  \\
AL-18x3  & 18   & 3    & 0.131 & 22.7      & 1054 & 0.796 & 0.316 & 1.038 & 1.104 & 1.90  \\
AL-9x3   & 9    & 3    & 0.131 & 23.8      & 1102 & 0.780 & 0.300 & 1.017 & 1.047 & 1.99  \\
AL-6x3   & 6    & 3    & 0.131 & 22.2      & 1030 & 0.767 & 0.285 & 1.000 & 0.996 & 2.19  \\
ST-18x3  & 18   & 3    & 0.131 & 26.9      & 1246 & 0.785 & 0.319 & 1.024 & 1.112 & 1.58  \\
ST-9x3   & 9    & 3    & 0.131 & 22.0      & 1020 & 0.787 & 0.320 & 1.027 & 1.116 & 1.64  \\
ST-6x3   & 6    & 3    & 0.131 & 23.3      & 1083 & 0.788 & 0.320 & 1.028 & 1.116 & 1.64 
\end{tabular}
\caption{Details of the LES cases with streamwise and spanwise spacing ($S_x/D$, $S_y/D$), local blockage ratio ($B$), simulated physical time in terms of flow-through times, number of revolutions, hydrodynamic coefficients ($C_T$ and $C_P$), their variation compared to the reference case AL-36x12, and resistance coefficient ($K$).} \label{table:setup}
  \end{center}
\end{table} 

\section{Large-eddy simulation results} \label{sec:results}

First, we focus on the array layouts with $S_x/D$ = 9 in order to elucidate how the flow field varies with $S_y/D$. The time-averaged and instantaneous flow fields are presented in \S\ref{sec:flowfield}, followed by some turbulence statistics in \S\ref{sec:turbulence}. We then present the wake centre-line velocities for all 28 cases in \S\ref{sec:centreline} to analyse the effects of both $S_x/D$ and $S_y/D$ on the wake recovery, and finally in \S\ref{sec:farmefficiency} the variations of the power and thrust coefficients are presented and compared with the theoretical predictions. Hereafter, the time-averaged value of any variable is denoted as $\langle \cdot \rangle$ and the instantaneous fluctuation value is represented as $(\cdot)'$ obtained from the Reynolds decomposition.
%, i.e., any instantaneous value of $u (t) = \langleu\rangle + u'(t)$ is divided into instantan. 

\subsection{Streamwise velocity field} \label{sec:flowfield}

\begin{figure}
\centerline{\includegraphics[width=.99\linewidth]{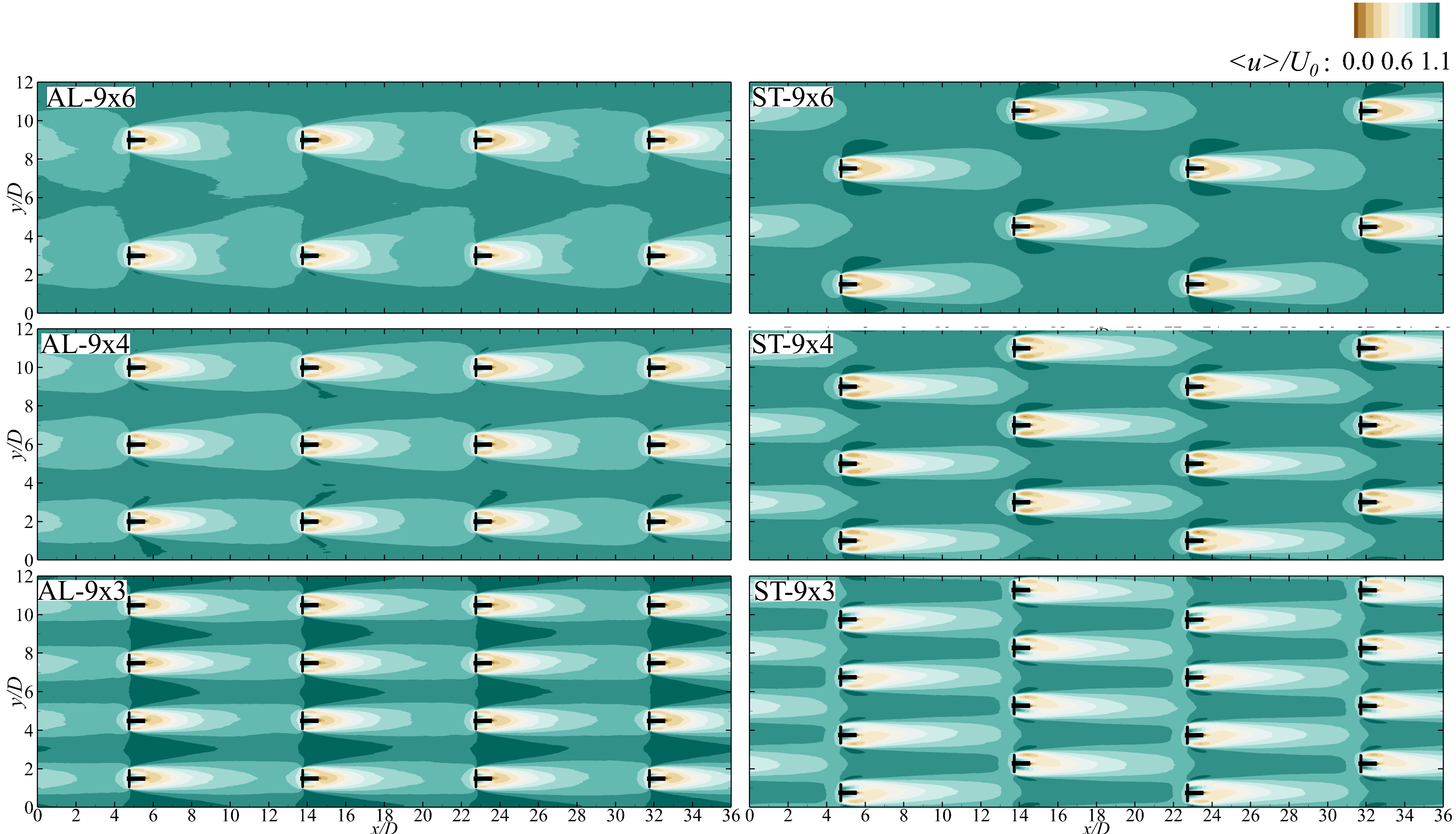}}
\caption{Contours of time-averaged streamwise velocity $\langle u \rangle/U_0$ at hub height for aligned and staggered cases with $S_x/D$ = 9 and lateral spacing $S_y/D$ = 6 (top), 4 (mid) and 3 (bottom).} \label{fig:les_cont_vel}
\end{figure}

We present in figure \ref{fig:les_cont_vel} contours of the normalised time-averaged streamwise velocity ($\langle u \rangle/U_0$) comparing aligned and staggered layouts with the same streamwise spacing of $S_x/D=9$ and lateral spacing of $S_y/D$ = 6, 4 and 3.
For aligned arrays, reducing the lateral separation between devices in the same row leads to an increased local blockage that induces larger flow acceleration in the bypass flow between them, which is well observed for AL-9x3 with the bypass-flow velocity being approximately 20\% higher than the bulk velocity. % \citep{Yang2012}. 
In perfectly staggered arrays, the wake generated behind each turbine is mostly recovered when reaching the following row at a downstream distance of $S_x$. 
Then, due to the lateral blockage caused by the turbines located on both sides, the mostly-recovered wake accelerates further and then impinges the turbine located in the next row, i.e. at 2$S_x$ downstream of the turbine that originally generated the wake. %which is beneficial to harness more energy as shown later in \S\ref{sec:farmefficiency}.
For layouts with low lateral blockage, i.e. $S_y/D$ = 6 and 4, there is a wider lateral spread of the wakes, which is well-observed in figure \ref{fig:les_cont_vel} for the aligned cases. 
However, in staggered configurations this lateral wake expansion appears limited compared to the corresponding aligned cases.

\begin{figure}
\centerline{\includegraphics[width=.99\linewidth]{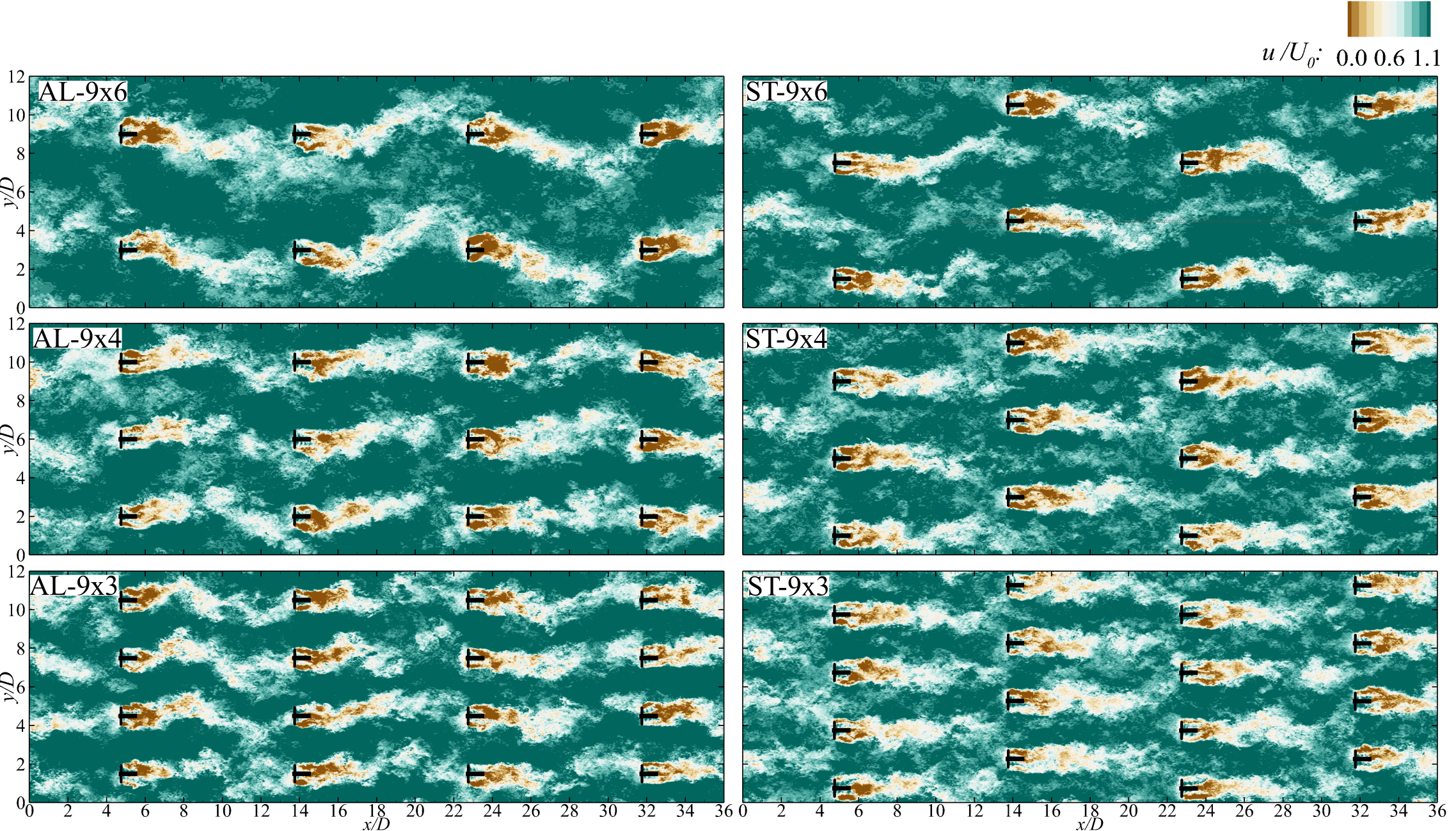}}
\caption{Contours of instantaneous streamwise velocity $u/U_0$ at hub height for aligned and staggered cases with $S_x/D$ = 9 and lateral spacing $S_y/D$ = 6 (top), 4 (mid) and 3 (bottom). Same legend as \ref{fig:les_cont_vel}.} \label{fig:les_cont_instvel}
\end{figure}

The different spreading rates of the time-averaged turbine wakes are partly explained by their instantaneous behaviour. 
In figure \ref{fig:les_cont_instvel} we present contours of instantaneous streamwise velocities ($u/U_0$) for the previous cases shown in figure \ref{fig:les_cont_vel} with $S_x/D$ = 9 and changing $S_y$, which reveals the pronounced meandering nature of the wakes with large oscillation amplitudes \citep{Foti2019}.
In aligned configurations, the meandering of the wake is affected by the lateral distance between turbines in the same row, i.e., increasing $S_y$ leads to a larger meandering amplitude. This is presumably because a higher lateral blockage can constrain the spanwise wake motion.
It can also be seen that there is a larger wake meandering amplitude in aligned arrays compared to their staggered counterparts. The lateral wake motion due to wake meandering seems almost negligible for ST-9x3, explaining the narrow time-averaged wakes shown earlier in figure \ref{fig:les_cont_vel}.
% the wakes behind the turbines do not appear to meander due to the highly blocked environment they operate in.
%Hence, the wake meandering phenomenon has larger implications for turbines in aligned configurations as they are directly subjected to undergo larger low-frequency oscillations whilst the turbine disposition in staggered layouts limits the impact of the wake meandering.

\subsection{Turbulence statistics} \label{sec:turbulence}

\begin{figure}
\centerline{\includegraphics[width=.99\linewidth]{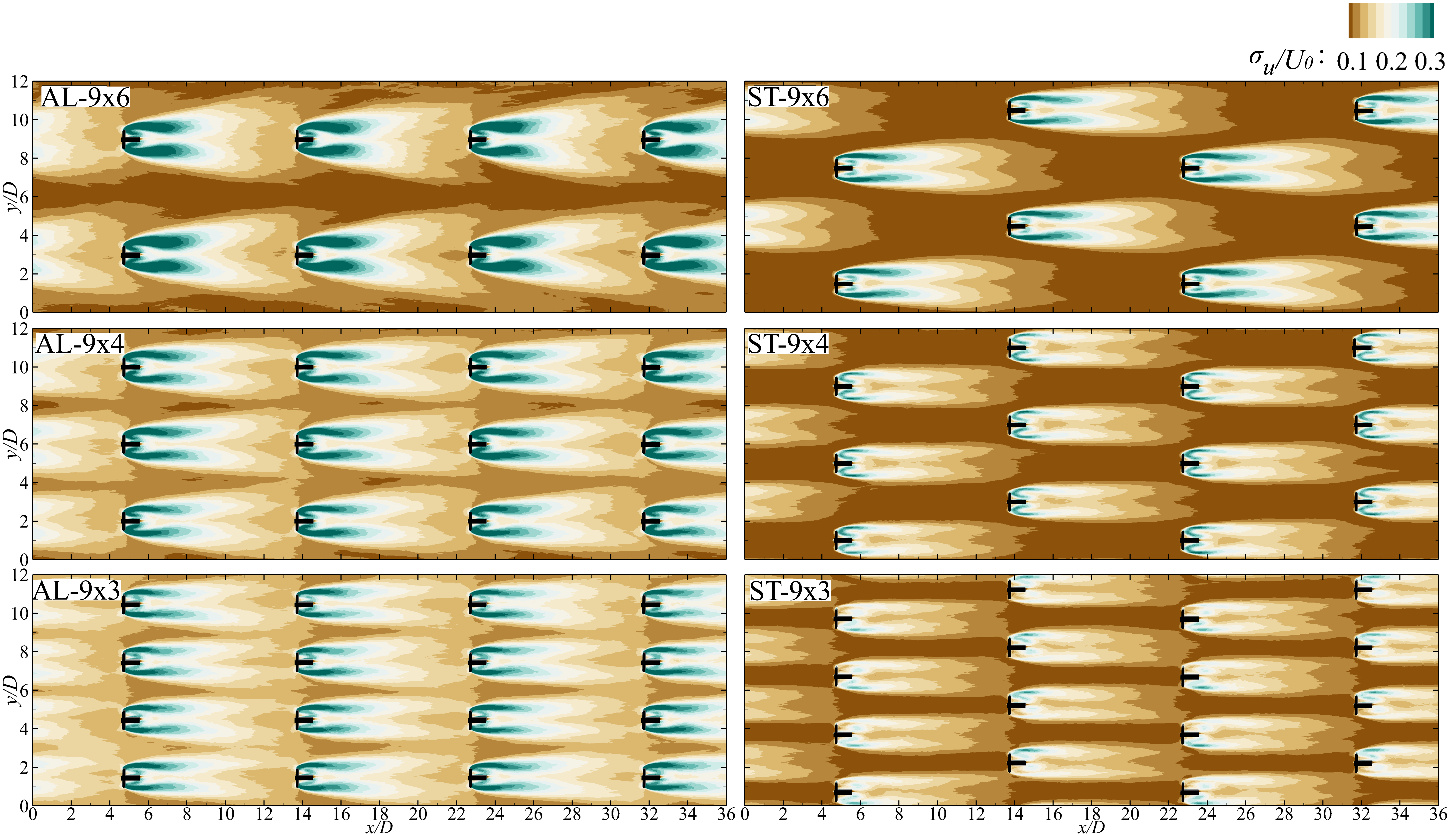}}
\caption{Contours of streamwise turbulence intensity ($\sigma_u/U_0$) at hub height for aligned and staggered cases with $S_x/D$ = 9 and lateral spacing $S_y/D$ = 6 (top), 4 (mid) and 3 (bottom).} \label{fig:les_cont_vel_up}
\end{figure}

\begin{figure}
\centerline{\includegraphics[width=.99\linewidth]{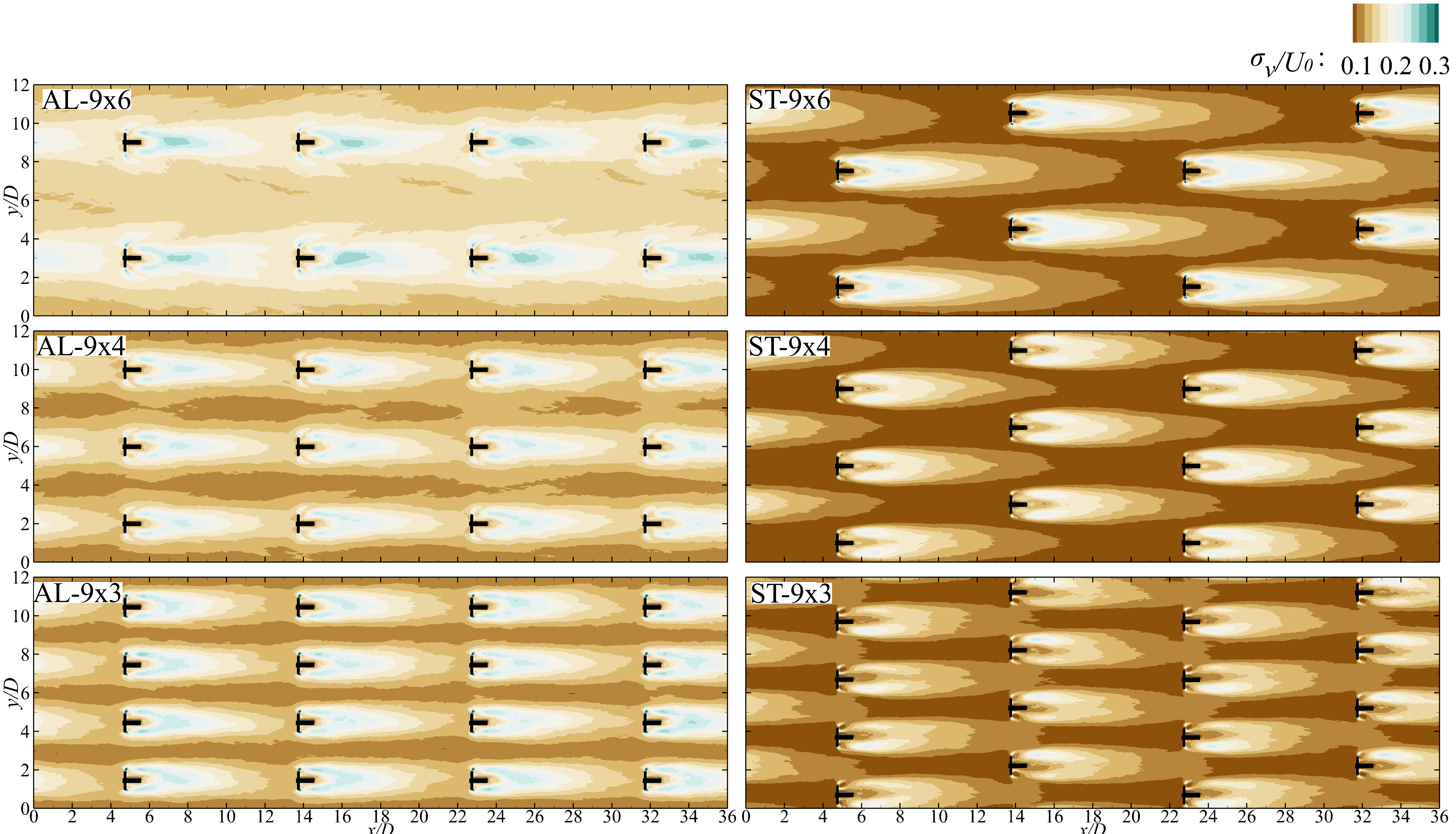}}
\caption{Contours of spanwise turbulence intensity ($\sigma_v/U_0$) at hub height for aligned and staggered cases with $S_x/D$ = 9 and lateral spacing $S_y/D$ = 6 (top), 4 (mid) and 3 (bottom).} \label{fig:les_cont_vel_vp}
\end{figure}

We further analyse the flow field in figures \ref{fig:les_cont_vel_up} and \ref{fig:les_cont_vel_vp} with contours of turbulence intensity in the streamwise ($\sigma_u/U_0$, with $\sigma_u = \langle u'u' \rangle^{0.5}$) and transverse directions ($\sigma_v/U_0$, with $\sigma_v = \langle v'v' \rangle^{0.5}$), again for the cases with $S_x/D$ = 9. 
These second-order statistics indicate that having turbines in an aligned layout leads to a notably stronger flow unsteadiness both inside and outside the wake of each turbine compared to the staggered cases. This agrees with the earlier observation that the wake meandering is stronger in the aligned cases.
In the near-wake region, in which tip vortices are still coherent and maintain a shear layer between the core flow and the bypass flow \citep{Ouro2019JFS}, both streamwise and spanwise turbulence intensities are substantially higher in the aligned cases than in the staggered cases.
% The area corresponding to the pathway described by tip vortices features higher values of $u'/U_0$ for aligned cases than staggered ones, linked to the ability of these turbulent structures to remain coherent for the former configurations. %%%%%[To Pablo: I'm not sure if your explanation is correct here - I presume that the high turbulence intensity in the aligned case is mainly due to the fact that the wake meandering is stronger, rather than due to the strength of tip vortices - note that the lateral motion of the entire wake region should increase not only v' but also u']%%%%%
Low turbulence intensity regions in the staggered cases are evident in the flow bypassing each turbine, coinciding with the regions in which the wake of an upstream turbine is almost fully recovered as shown earlier in figures \ref{fig:les_cont_vel} and \ref{fig:les_cont_instvel}.
The results also show that increasing the blockage ratio $B$ (or reducing the distance $S_y$ between turbines in the same row) reduces the turbulence intensity irrespective of the overall turbine arrangement, as a consequence of constraining the meandering motion. 

%\begin{figure}
%\centerline{\includegraphics[width=.9\linewidth]{Figures/Fig_AL_ST_LES_tke.pdf}}
%\caption{Contours of turbulent kinetic energy $k/U_0^2$ at hub height for aligned and staggered cases with $S_x/D$ = 9 and lateral spacing $S_y/D$ = 6, 4 and 3. \textbf{PABLO: perhaps remove this one?}} \label{fig:les_cont_vel_tke}
%\end{figure}

\begin{figure}
\centerline{\includegraphics[width=.99\linewidth]{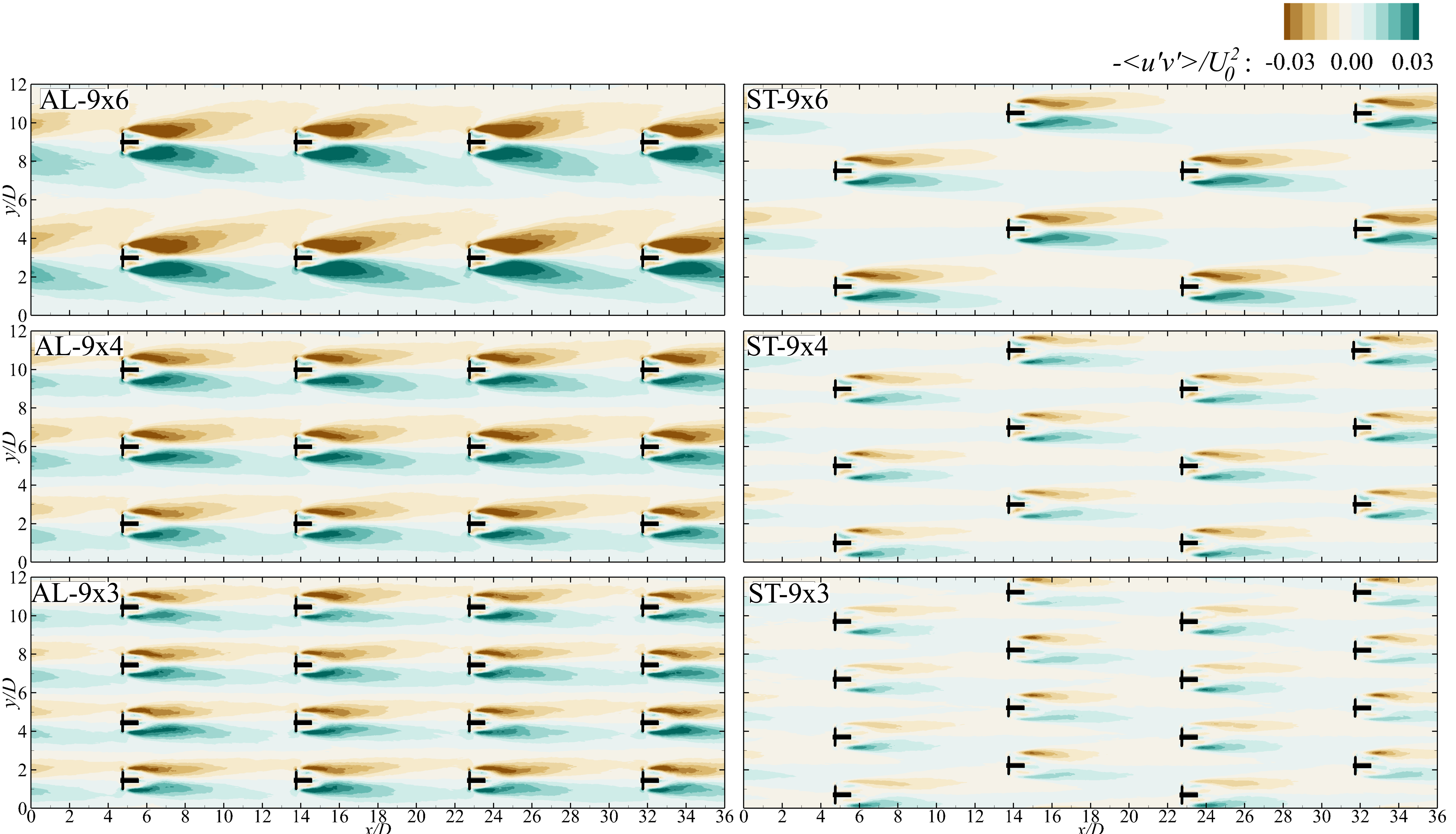}}
\caption{Contours of Reynolds shear stress ($-\langle u'v' \rangle/U_0^2$) obtained at hub height for aligned and staggered cases with $S_x/D$ = 9 and lateral spacing $S_y/D$ = 6 (top), 4 (mid) and 3 (bottom).} \label{fig:les_cont_vel_upvp}
\end{figure}

As can be expected from the above results, the wake meandering also notably affects the distribution of turbulent momentum exchange, which is described in figure \ref{fig:les_cont_vel_upvp} with contours of the Reynolds shear stress $-\langle u'v' \rangle/U_0^2$ at hub height. For each array layout, narrowing the lateral spacing reduces the level of Reynolds shear stress, indicating that there is a lower level of momentum exchange between the wakes and the surrounding bypass flows. Staggered layouts consistently attain a lower shear stress level compared to the aligned layouts irrespective of the number of turbines deployed.

%%% %\newpage \vspace{20cm} \pagebreak

\subsection{Centre-line velocities} \label{sec:centreline}

\begin{figure}
\centerline{\includegraphics[width=.99\linewidth]{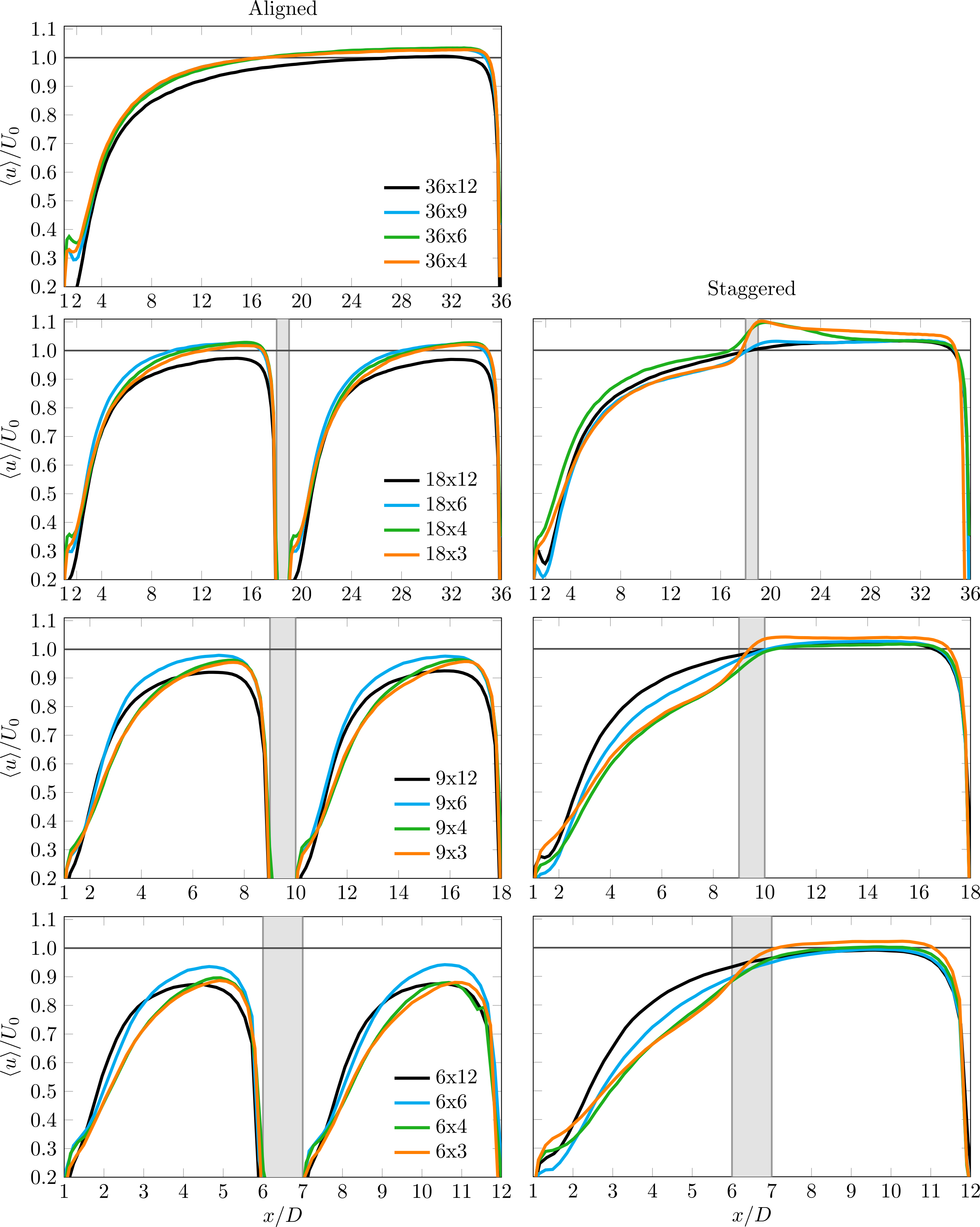}}
\caption{Distribution of time-averaged streamwise velocities ($\langle u\rangle/U_0$) along the wake centre-line at the hub height for aligned (left) and staggered (right) configurations with $S_x/D$ = 36 (top), 18 (mid-top), 9 (mid-bottom) and 6 (bottom), and $S_y/D$ = 12 (black), 6 (blue), 4 (green) and 3 (orange).} \label{fig:les_Ucentreline}
\end{figure}

Next, we quantitatively compare the mean streamwise velocity ($\langle u \rangle /U_0$) distribution along the wake centre-line in figure \ref{fig:les_Ucentreline} for every configuration simulated (see table \ref{table:setup}).
Note that for $S_x/D$ = 18, 9 and 6 we plot the velocity distribution over two rows of turbines, including a shaded area representing the location at which turbines from the second row are located.
When the streamwise spacing is relatively small ($S_x/D$ = 9 and 6), the values of $\langle u \rangle/U_0$ behind the turbine nacelles increase faster for the lower lateral blockage cases, i.e. $S_y/D$ = 6 and 12.
This faster wake recovery in lower blockage cases results from a larger entrainment of ambient flow into the wake enhanced by the stronger meandering behaviour, as seen earlier in the distribution of Reynolds shear stresses in figure \ref{fig:les_cont_vel_upvp}.

Although lower blockage ratios (or larger $S_y/D$ values) tend to result in higher wake recovery rates in the near-wake region, this trend does not hold in the far-wake region. In the aligned cases with $S_x/D$ = 18, the values of $\langle u\rangle/U_0$ are slightly over unity shortly upstream of the turbines for configurations with $S_y/D \leq $ 6, whereas for $S_y/D$ = 12 the wake velocity is not fully recovered to the bulk velocity value despite the high wake recovery rate in the near-wake region.
Further decreasing $S_x/D$ to 9 in aligned cases reduces the amount of wake recovery for every $S_y/D$, resulting in $\langle u\rangle/U_0$ of about 0.9 to 0.95 at one rotor-diameter upstream of the turbines ($x/D$ = 8 and 17). It is clearly seen that the wakes in higher blockage cases (AL-9x4 and AL-9x3) have a lower recovery rate in the near-wake region but a higher recovery rate in the far-wake region.
The same trend can be observed in the smallest streamwise spacing cases ($S_x/D$ = 6), where the streamwise velocities upstream of the turbines are even lower due to the lack of space for the wake to recover.
Overall, for aligned arrays we may conclude that $S_x/D$ is the main design parameter that determines how much the wake recovers before approaching the turbine downstream, with $S_y/D$ contributing to a lower extent by changing the turbulent wake characteristics.

In staggered arrays with $S_x/D$ = 18, the centre-line mean velocity notably increases at around $x/D$ = 18, which coincides with the location of the second row shifted laterally with respect to the first row of turbines.
This flow acceleration is due to the local blockage caused by the turbines in the second row, and thus is more noticeable when the lateral spacing $S_y/D$ is small. For example, for ST-18x3 and ST-18x4 the maximum velocity is about 1.1$U_0$, whereas for ST-18x6 or ST-18x12 it is about 1.02$U_0$. This agrees qualitatively with the theoretical analysis presented earlier in \S\ref{sub:theory_e}, i.e., the local flow velocity upstream of each turbine may exceed the cross-sectional average velocity in staggered arrays.
Reducing the streamwise spacing $S_x/D$ makes this additional local flow acceleration less noticeable, but the rate at which the wake velocity recovers still varies with $S_y/D$, being higher for the higher blockage cases with $S_y/D$ = 4 and 3.
Another distinct feature of time-averaged wakes in staggered arrays is that after passing through the streamwise location of the second row (i.e., $x \geq S_x$) the centre-line mean velocity remains fairly constant until about two rotor-diameters upstream of the next turbine.

Finally, comparing aligned and staggered cases for a given streamwise spacing, we can confirm that the velocity recovery rate in the near-wake region is higher in the former configurations than in the latter. This is due to stronger wake mixing enhanced by the wake meandering as shown earlier.

%%% 
%\newpage \vspace{20cm} \pagebreak

\subsection{Array efficiency} \label{sec:farmefficiency}

\begin{figure}
\centerline{\includegraphics[width=.9\linewidth]{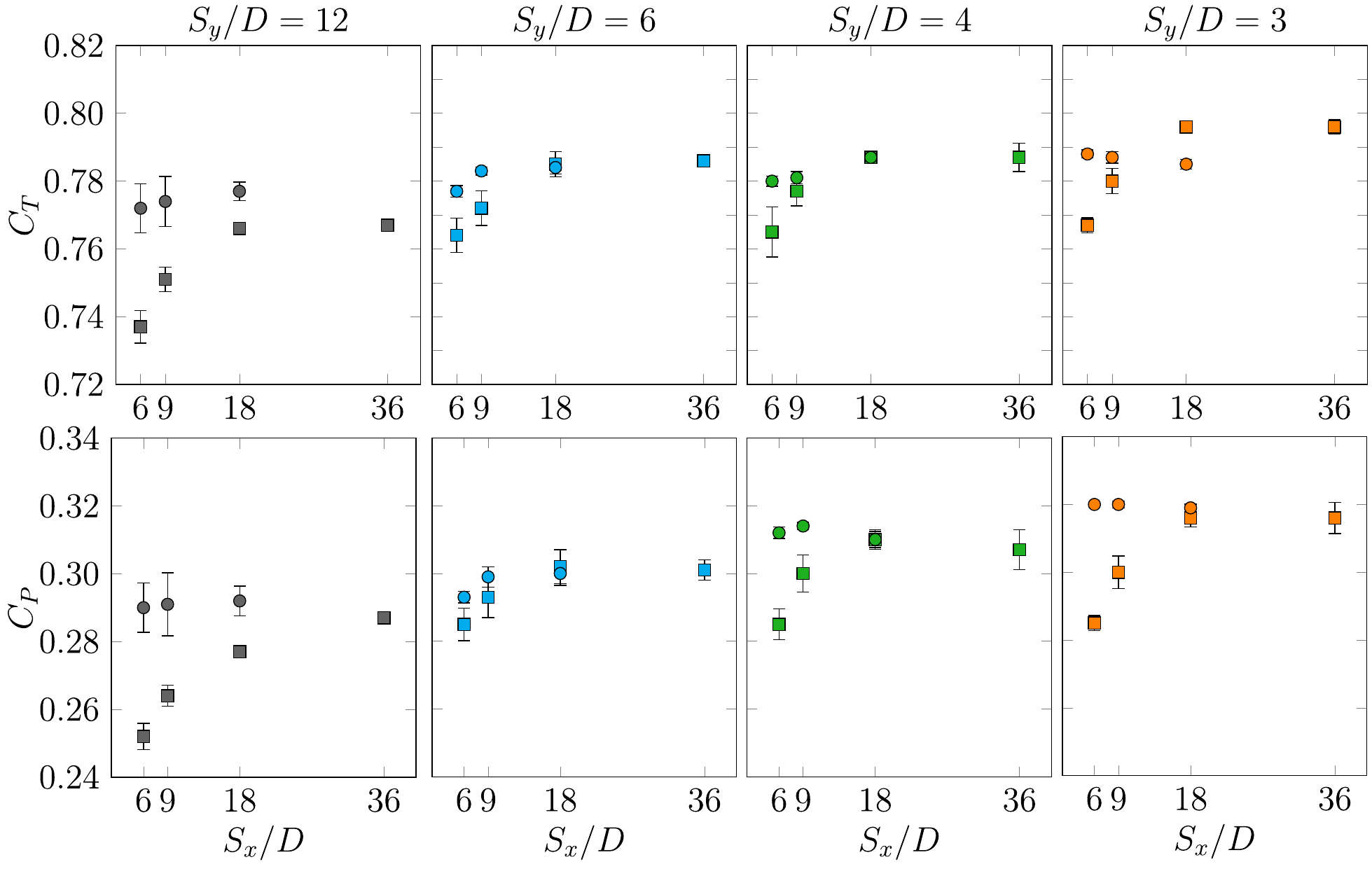}}
\caption{Results of $C_T$ (top) and $C_P$ (bottom) obtained from the LES for the aligned ($\square$) and staggered ($\bigcirc$) layouts.} \label{fig:les_cxcp}
\end{figure}

As the impact of array layout and turbine spacing on the flow field has been confirmed, we now focus on the thrust and power coefficients of turbines to analyse their efficiency in the simulated infinitely large tidal arrays.
For each configuration, time-averaged hydrodynamic coefficients are further averaged for all turbines comprising the array, and the results are presented in figure \ref{fig:les_cxcp} with error bars indicating the standard deviation of array-averaged values.
For aligned layouts with a given $S_y/D$, both thrust and power coefficients tend to have their maximum values at $S_x/D$ = 36, but the results at $S_x/D$ = 18 are similar as this streamwise spacing is still large enough for the wakes to almost fully recover. However, when $S_x/D$ is further reduced, both $C_T$ and $C_P$ drop considerably in line with the decrease in the mean streamwise velocity upstream of each turbine, as shown earlier in figure \ref{fig:les_Ucentreline}.
In contrast, the performance of turbines in staggered configurations appears less sensitive to $S_x/D$.

Comparing arrays with large lateral spacing ($S_y/D$ = 12), staggered configurations consistently provide higher $C_T$ and $C_P$ values than the aligned counterparts, with larger differences for shorter streamwise spacing. 
For cases with $S_y/D$ = 6 and 4, staggered configurations still tend to provide higher $C_T$ and $C_P$ than the aligned counterparts, although the differences are negligibly small at $S_x/D$ = 18. Further reducing the lateral spacing to $S_y/D$ = 3 leads to higher $C_T$ in the aligned case than in the staggered case at $S_x/D$ = 18, but again $C_P$ values are consistently higher in the staggered cases. These results suggest that, as expected, staggered configurations are in general more efficient than the aligned counterparts.
%The highest $C_P$ value (of $\approx$ 0.320) is obtained for the staggered case with $S_y/D$ = 3 and almost irrespective of $S_x$, meaning that the blockage effect is not too relevant for the performance of these staggered arrays, as previously observed in figure \ref{fig:les_Ucentreline}.
%However, for the aligned case the peak $C_P$ is 0.316 for $S_x/D=18$ and $S_y/D=3$ and its value decreases when narrowing the distance between rows, a consequence of the reduced kinetic energy recovery in the wake (figure \ref{fig:les_Ucentreline}). 

%Tidal turbine arrays with $S_y/D=3$ experience a large lateral blockage that can disturb the flow around the turbines (figure \ref{fig:les_cont_vel}.
%These LES results suggest that aligned arrays with larger streamwise spacing reach larger thrust and power coefficients, which are reduced when increasing $S_x/D$.
%Staggered array with $S_x/D$ = 18 and $S_y/D$ = 3 obtains the largest $C_P$ amongst the analysed cases as well as the highest thrust-to-power coefficients ratio of 0.431.

%These results suggest that large spacings enable a larger momentum recovery from the wakes and thus enable higher turbine performance, except for the staggered cases with low lateral blockage ($S_y/D=6)$.

\begin{figure}
\centerline{\includegraphics[width=.95\linewidth]{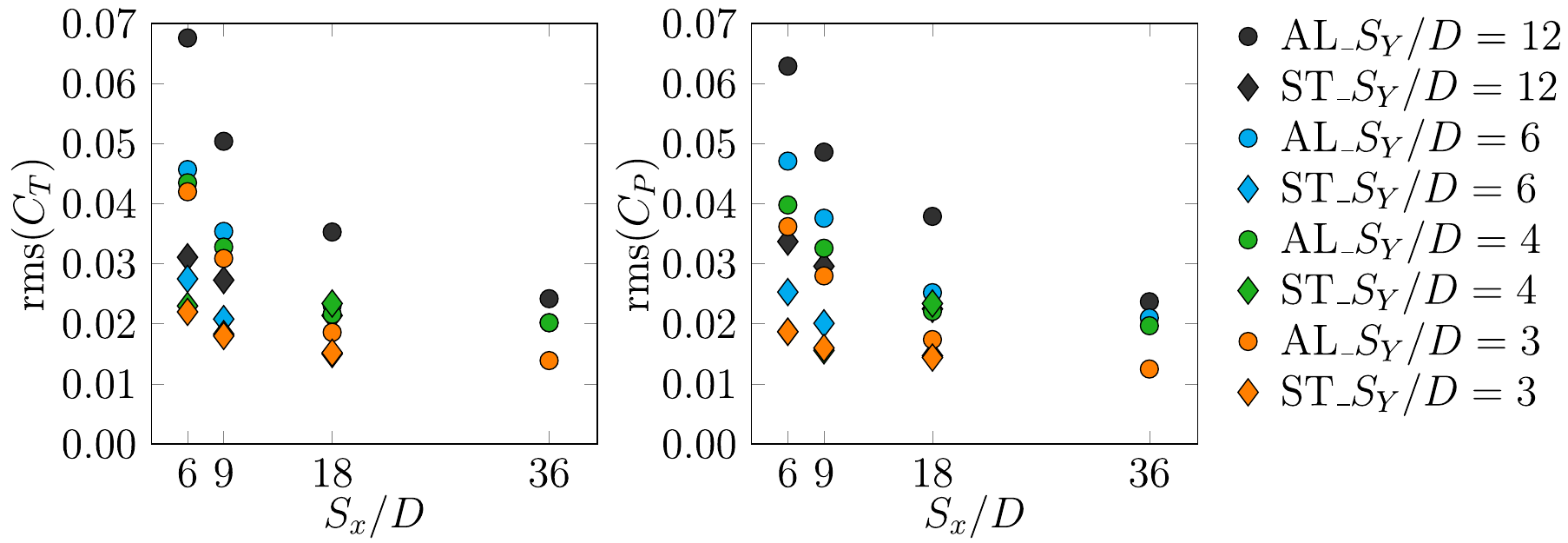}}
\caption{Results of root-mean-squared temporal fluctuations of $C_T$ (left) and $C_P$ (right) obtained from the LES of the considered array layouts.} \label{fig:les_rmscxcp}
\end{figure}

%The unsteadiness of the approaching flow is a result of the flow dynamics introduced by upstream devices, e.g. enhanced turbulence levels and wake meandering, which lead to force fluctuations that are of importance for long-term fatigue loads.
Next, we analyse the temporal fluctuations of thrust and power from their root-mean-square (rms) values. The results averaged over all turbines comprising the array are presented in figure \ref{fig:les_rmscxcp}, which shows that decreasing $S_x/D$ leads to an increase in the fluctuations of both thrust and power, as expected from enhanced unsteadiness of the approaching flow. 
Adopting a large lateral spacing also increases the load fluctuations in comparison to cases with the same streamwise spacing and a smaller lateral spacing. This is in line with the strength of wake oscillations increasing with $S_y/D$, as shown earlier in figure \ref{fig:les_cont_instvel}.
It is also worth noting that turbines in aligned configurations tend to have larger load variations than the staggered counterparts, especially when decreasing $S_x/D$.
These results suggest that staggered configurations not only enhance the array's efficiency (i.e., mean power coefficient) but can also reduce the load fluctuations and thus the long-term fatigue damage of turbine rotors.

%In our simulations of infinitely long tidal arrays, we can quantify the drag resistance imposed by the turbines in the flow with the friction velocity derived from the pressure gradient that keeps a uniform flow rate. 
%This friction velocity accounts for both the bottom roughness shear and turbine thrust. 
%We plot the change of $u^*$ with streamwise spacing in figure \ref{fig:comparison_CP}.
%Comparing aligned and staggered arrangements for the same layout, LES results indicate that the overall drag resistance is similar, with the largest deviation observed for the case when $C_P$ reaches its maximum, i.e. $S_x/D$ = 9 and $S_y/D$ = 1, being $u^*$ reduced about 12\% when turbines are staggered. 

%Further insights into the tidal array effective power and turbine efficiency are presented in figure XXX with results in terms of the array's effective power (EP) that accounts for the total power generated compared to the case with a single turbine, i.e. AL\_36x18 for our case, and characteristic efficiency ($\eta$) of the turbines, which is the ratio between the overall power production for each case and that produced by the same number of turbines using the power generated in the single turbine case.

%%% 
%\newpage\pagebreak
\section{Discussion}\label{sec:discussion}

The layout of turbines in a large tidal array has important implications to its power generation capability. This is because the flow through a large array is characterised by complex wake-turbine interactions involving the effects of local blockage and wake mixing, both of which are functions of the layout of turbines.
In this study we aimed to understand the impact of turbine layout for infinitely large tidal arrays, using a new theoretical model based on the LMADT and high-fidelity LES-ALM, focusing on how the array efficiency changes depending on the turbine resistance, local blockage ratio within each row of turbines ($B$ in the theoretical analysis, which is a function of $S_y$ in the LES, i.e. $B \propto S_y$) and the completeness of wake mixing between each row ($m$ in the theoretical analysis, which is a function of $S_x$ in the LES, i.e. $m \propto S_x$).
It should be borne in mind that the theoretical model allows to predict overall trends and upper bound estimates of the array efficiency with almost negligible computational cost, whereas the LES-ALM allows to resolve the details of complex turbulent flow field within the array but at a high computational expense. Thus, these are two highly contrasting approaches to this fluid flow problem.

Before comparing and further discussing the theoretical and LES-ALM results, we emphasise two key differences between these two approaches: 
(i) In the LES-ALM the operating point of turbines (i.e., rotational speed) is kept constant for all cases, resulting in slightly different $K$ values as presented in table \ref{table:setup}. No further tuning of the rotational speed to obtain a constant $K$ (to match the theoretical analysis) is performed due to the high computational cost that would be required for it.
Alternatively, a fairer comparison with the theoretical analysis could be made using LES with an Actuator Disc Model (ADM), but we adopted an ALM in this study as it allows a more realistic representation of the turbulent flow field within the array. 
Thus, consideration needs to be taken when comparing the theoretical results for a constant $K$ value with scattered LES results with slightly different $K$ values.
(ii) The current theoretical model does not provide an explicit relationship between the mixing factor $m$ and the array configuration, whereas the LES-ALM automatically predicts the wake recovery rate for a given configuration by resolving the turbulent flow field. To make a direct comparison, here the mixing factor in the theoretical analysis is assumed to be $m = 1 - (S_x/D)^{-1}$, which is arguably the simplest model to relate $m$ to $S_x/D$ without knowing any detailed characteristics of turbine wake mixing for a given array configuration a priori.

%%%(iii) LMADT assumes the wake region to be divided into viscous and inviscid zones, whilst LES shows that the flow is turbulent (i.e., viscous forces are present) over the whole wake length. Nevertheless, the presence of an inviscid wake region appears to be somehow true in staggered layouts as once the intermediate turbine row location is surpassed streamwise velocities remain nearly constant. Thus, our time-averaged LES results indicate the inviscid-viscous assumption can hold true.

\begin{figure}
\centerline{\includegraphics[width=.95\linewidth]{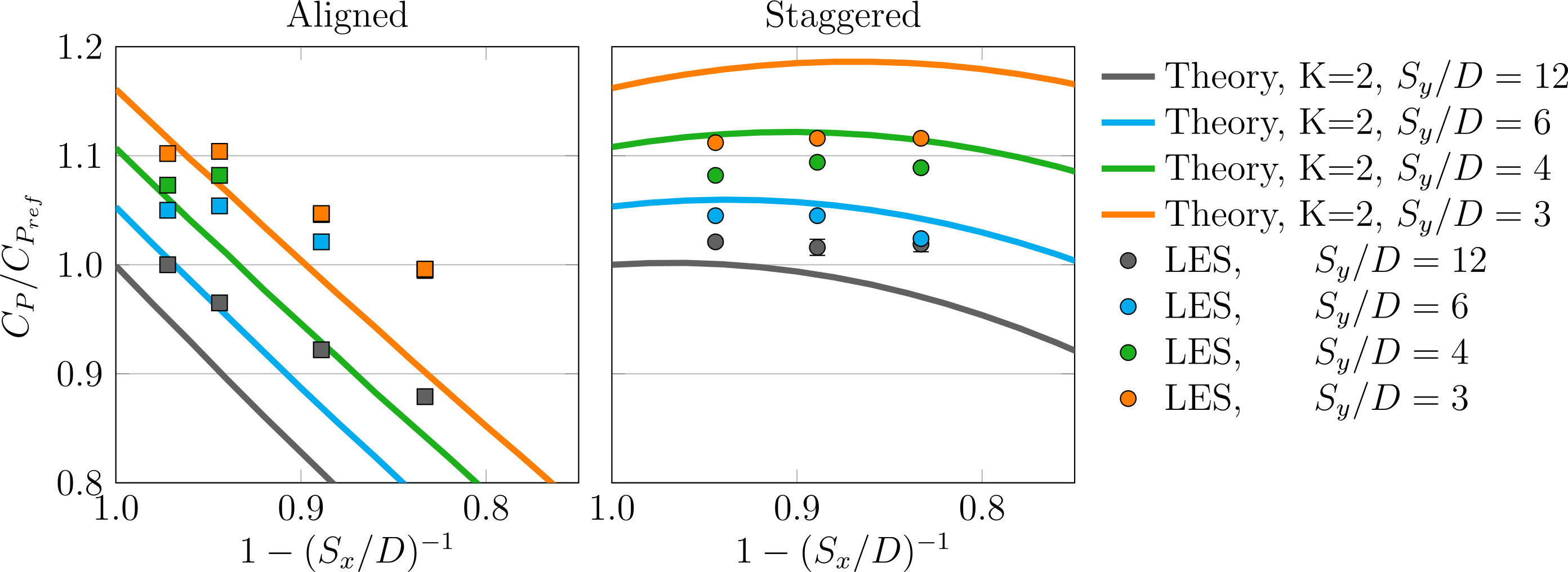}}
\caption{Comparison of normalised power coefficient ($C_P$/$C_{P_{ref}}$) predicted with the theory (assuming $K=2$ and $m = 1 - (S_x/D)^{-1}$) and computed from the LES.} \label{fig:comparison_CP_LxD}
\end{figure}

We now compare array efficiency predictions between the theoretical analysis (assuming $K$ = 2) and LES-ALM in figure \ref{fig:comparison_CP_LxD}. Since the theoretical $C_P$ values are for ideal turbines (or actuator discs) and they are not directly comparable to $C_P$ values for real rotors, here we normalise $C_P$ with a reference power coefficient $C_{P_{ref}}$, which is defined for the theoretical analysis and LES-ALM separately. For the LES-ALM, $C_{P_{ref}}$ is the power coefficient obtained for the sparsest array (AL-36x12), in which turbine-to-turbine interactions are deemed negligible, whereas for the theoretical analysis, this is the power coefficient for $B=0.033$, $K=2$ and $m=1$. Hence, $C_P/C_{P_{ref}}$ plotted in figure \ref{fig:comparison_CP_LxD} represents the change rate of $C_P$ due to the effect of different turbine layouts.
Overall, there is a qualitative agreement in $C_P/C_{P_{ref}}$ between the two approaches. In aligned arrays, the efficiency monotonically decreases with $S_x/D$ (except when $1 - (S_x/D)^{-1}$ is close to unity and $S_y/D \leq$ 6) as turbines increasingly operate in the wake of upstream turbines. 
The rate of decrease in $C_P/C_{P_{ref}}$ is different between the theoretical and LES results, largely due to the simple relationship between $m$ and $S_x/D$ assumed to make this comparison. It is therefore expected that the agreement would improve if the relationship between $m$ and $S_x/D$ is modelled appropriately in future work. For staggered arrays, again the theoretical predictions agree qualitatively with the LES, showing that $C_P/C_{P_{ref}}$ is insensitive to the streamwise spacing at least within the range of conditions considered here. However, the effect of lateral spacing $S_y/D$ is slightly over-predicted by the theoretical model compared to the LES.

%%%Previous efficiency analyses with LMADT have suggested that in confined tidal channel flows the array layout that achieves maximum performance is deploying turbines in a single row (or fence) \citep{Nishino2012,Nishino2013,Draper2014} as it neglects wake-turbine effects and enhances local blockage.
%%%In this study we consider unconfined, infinitely long tidal arrays and thus some differences in the observations can be expected as adopting periodic streamwise conditions means that for all LES configurations turbines experience wake effects.
%%%We observe that LES-computed $C_P$ values obtained for single-row arrays (i.e., $S_x/D$ = 36) are not consistently the highest for all blockages as staggered arrays yielded larger values for $S_y/D$ = 12 or 4, as seen in figures \ref{fig:les_cxcp} and \ref{fig:comparison_CP_LxD} and table \ref{table:setup}.

The above comparison of $C_P/C_{P_{ref}}$ suggests that the new theoretical tidal array model is promising as it seems to capture the combined effects of local blockage and wake mixing qualitatively correctly, despite not accounting for some key transient flow phenomena, such as blade tip vortices and wake meandering, which are well-captured in the LES-ALM. 
As the efficiency of turbines in a large array is determined mainly by the time-averaged flow field within the array, further improvements of the theoretical model could be made in future studies using LES. In particular, further results of LES-ALM for a wider range of parameters would allow us to empirically model the mixing factor $m$ as a function of both $S_x/D$ and $S_y/D$, where it would be important to account for the dependency of wake meandering and other transient flow phenomena (that collectively determine the wake recovery rate) on the layout of turbines. It should be remembered, however, that the efficiency of real tidal arrays would depend not only on micro-scale flow interactions within the array, namely the turbine-to-turbine interactions studied here, but also on macro-scale flow interactions outside the array \citep{Vennell2012,Vennell2015,Gupta2017}.

%%% %%%%%%%%%%%%%%%%
%\newpage\pagebreak
\section{Conclusions}\label{sec:conclusion}

This paper has investigated the performance of tidal stream turbines in an infinitely large array, using two different approaches: a quasi-one-dimensional theoretical model based on the Linear Momentum Actuator Disc Theory (LMADT), and Large-Eddy Simulation with an Actuator Line Method (LES-ALM). Two different types of turbine layouts were considered in this study, i.e., perfectly aligned and staggered layouts. For the LES-ALM, 28 different arrays with various streamwise ($6 \leq S_x/D \leq 36$) and lateral ($3 \leq S_y/D \leq 12$) turbine spacing were considered. For the theoretical analysis, a hybrid inviscid-viscous approach was employed to model an infinitely large array using only three input parameters, namely the local blockage ratio $B$, disc resistance coefficient $K$ and wake mixing factor $m$.

Our LES results have shown that the lateral spacing has a pronounced effect on the characteristics of wake meandering. In particular, the amplitude of wake meandering is found to decrease as $S_y/D$ is reduced. 
The main consequence of this change in wake dynamics is that a lower amplitude of wake meandering leads to less entrainment of momentum from the surrounding bypass flow into the wake, and thus a lower wake recovery rate. However, this negative effect of small lateral spacing on the wake recovery rate is observed mainly in the near-wake region only, and the completeness of wake recovery (i.e., how much the wake velocity is recovered before the wake approaches the next turbine) tends to depend more on the streamwise spacing, especially for aligned arrays. We have also confirmed from our LES results that, in staggered arrays, the wake experiences an additional acceleration when it passes through the laterally-shifted row of turbines immediately downstream. This additional acceleration is due to the effect of local blockage, which is enhanced when the lateral spacing is small. When $S_y/D$ is sufficiently small, the centre-line wake velocity is found to even exceed the bulk velocity, resulting in a high power of turbines for a fixed bulk velocity.
Resolving the turbulent flow field with LES has also allowed to study the temporal fluctuations of turbine loads in a large tidal array, showing that these are approximately twice larger in aligned arrays than in staggered arrays for a given turbine spacing.

%%%Hence, from a long-term operative perspective, turbines in aligned arrays ought to suffer larger fatigue loads than devices deployed in an staggered fashion.

%%%In these laterally unconfined flows, both lateral blockage $B$ between turbines in the same row and rate of wake mixing (i.e., how much kinetic energy is recovered in the wake region) are the key phenomena driving array performance ($C_P$).
%%%Mean velocities computed from LES show that momentum in the wake is recovered before impinging turbines in all staggered layouts but in aligned cases the available kinetic energy to be harnessed diminishes when decreasing $S_x$.

Whilst the LES results have revealed the complexity of turbulent flow phenomena that collectively determine the performance of turbines in a large tidal array, the simple theoretical model has captured the basic trend of turbine performance in both aligned and staggered arrays qualitatively correctly. In particular, the theoretical model suggests that there is an optimal streamwise spacing to maximise the performance of turbines (or the power of turbines for a fixed bulk velocity) in a large staggered array. This optimum exists as the local flow velocity upstream of each turbine can exceed the bulk velocity only when the streamwise spacing is reasonably (not excessively) large for the mixing between locally faster and slower flows to be largely (but not entirely) completed within that streamwise distance. However, both theoretical and LES results show that, at least within the range of conditions tested, the effect of streamwise spacing is less than that of lateral spacing, i.e., the performance of turbines in staggered arrays depends more on $S_y/D$ than on $S_x/D$. We have also observed some quantitative differences between the theoretical predictions and the LES. One of the main causes of differences is that, to make a comparison between the two approaches, we have assumed a simple relationship between the mixing factor $m$ (representing the completeness of mixing after each row of turbines) and the streamwise spacing between rows. Further LES results for a wider range of parameters would be helpful in future studies to develop an empirical model of $m$ as a function of both streamwise and lateral spacing.

The results obtained in this study will help understand and improve the performance of tidal turbines in future large tidal arrays. It should be borne in mind, however, that the performance of real tidal arrays may depend not only on micro-scale (turbine-to-turbine) flow interactions within the array but also on macro-scale flow interactions between the array and tidal-channel flow, the latter of which was outside the scope of this study. 

%%%%%%%%%%%%%
\section*{Declaration of interests}
The authors report no conflict of interest

%%%%%%%%%%%%%
\section*{Acknowledgements}

This research has been partially funded by the UK's Engineering and Physical Sciences Research Council (EPSRC) (grant number EP/R51150X/1).
The first author would like to acknowledge the support of the Supercomputing Wales project, which is partially funded by the European Regional Development Fund (ERDF) via the Welsh Government, the Isambard project funded by the EPSRC (EP/P020224/1), the GW4 alliance, the Met Office, Cray and Arm.
This work also used the ARCHER UK National Supercomputing Service (http://www.archer.ac.uk).
The second author would like to thank Dr Scott Draper for useful discussions on LMADT.

\section*{Appendix A: Large-eddy simulation code DOFAS}\label{app:DOFAS}

We used the Digital Offshore Farms Simulator (DOFAS) \citep{Ouro2019JFS}, an in-house LES code fully parallelised with Message Passing Interface (MPI) that also features a hybrid MPI/OpenMP scheme to maximise its computational performance \citep{Ouro2019CAF}. 
In DOFAS, the spatial domain is divided into rectangular sub-domains and discretised using Cartesian grids with staggered storage of velocities, i.e. velocity components are computed at the cell faces whilst pressure and scalar values are calculated at the cell centres. 
This scheme allows an even subdivision of the computational region into sub-domains to effectively perform the simulations. 
The governing equations resolved in DOFAS are the incompressible spatially-filtered Navier-Stokes equations:
\begin{eqnarray}
&& \frac{\partial u_i}{\partial x_i} = 0, \\ 
&& \frac{\partial u_i}{\partial t} + u_j \frac{\partial u_i}{\partial x_j}  = 
-\frac{1}{\rho} \frac{\partial p}{\partial x_i} 
+ (\nu + \nu_t) \frac{\partial^2 u_i}{\partial x_j^2} + f_t + \Pi_i, \label{eq:ns2}
\end{eqnarray}
where $u_i$ = $(u,v,w)^T$ is the vector of spatially-filtered velocities, the coordinates vector is $x_i$ = $(x,y,z)^T$, $\rho$ denotes the fluid density, $p$ is the relative pressure, $\nu$ is the kinematic viscosity of the fluid, and $f_t$ is a source term resulting from the actuator line method and immersed boundary forcing used for the representation of turbine rotors and nacelles, respectively.
$\Pi_i$ is a source term representing the driving pressure gradient responsible for keeping a constant flow rate when periodic boundary conditions are used in the streamwise direction.
%%%The latter term is used to account for the friction velocity $u_*^2$ in the streamwise direction as a representation of the total drag resistance originated from the bottom wall shear stress, $\tau_w = \Pi_1 H/\rho$ with $H$ denoting the water depth, and turbines rotor forces introduced by the actuator line and immersed boundary methods. 
The eddy-viscosity $\nu_t$ is calculated using the Wall-Adapting Local Eddy-viscosity (WALE) sub-grid scale model from \citet{Nicoud1999}. %%%as,

%%%\begin{equation}
%%%    \nu_t = ( C_w \Delta x)^2 \frac{ \left( S^d_{ij} S^d_{ij} \right)^{3/2} }{
%%%    \left( \overline{S}_{ij} \overline{S}_{ij} \right)^{5/2}  + \left( S^d_{ij} S^d_{ij} \right)^{5/4}     }
%%%\end{equation}

%%%with $C_w$ being a constant equal to 0.46, $\Delta x = x_i-x_{i-1}$ is the grid resolution, $S^d_{ij}$ is the traceless symmetric part of the square of the velocity gradient tensor and $\overline{S}_{ij} = 0.5 (\partial_j u_i + \partial_i u_j)$ is the strain-rate tensor of velocities.

The velocity field is spatially discretised using a fourth-order central differences scheme.
Simulations are advanced in time using a fractional step method, with a three-step low-storage Runge-Kutta scheme to obtain the non-solenoidal velocity field by explicitly computing the convection and diffusion terms, which is then corrected after the Poisson pressure equation is solved using an efficient multigrid solver \citep{Cevheri2016}. 

For the representation of solid bodies, DOFAS adopts a discrete direct-forcing Immersed Boundary Method (IBM) with pointwise interpolating delta functions, which has been validated in studies including tidal turbines \citep{Ouro2017JFS,Ouro2018FTC}, geophysical flows \citep{Ouro2017POF}, rough open-channel flows \citep{Stoesser2010,Bomminayuni2011,Nikora2019JFM}, and fluid-structure interaction \citep{Kara2015b,Ouro2019PRF}. In the present work, the IBM is adopted to represent the turbine nacelles, using the $\phi_4$ delta function for the interpolation procedures. 
Turbine rotors are represented using an Actuator Line Model (ALM) validated in \citet{Ouro2019JFS} which discretises the blades into a set of $N_L$ points, evenly spaced as a function of the mesh resolution. 
The ALM has been proven to provide an adequate description of the wake dynamics for wind and tidal turbines \citep{Breton2017}.

In this study we set turbine rotors to have a constant rotational speed, $\Omega$, and use prescribed lift and drag coefficients of hydrofoils tabulated for a range of angles-of-attack to obtain the lift and drag forces at every point comprising the turbine blades. From the drag and lift forces we calculate the thrust force $T$ and tangential force $Q$, and thus determine the generated power $P=Q \Omega$. Eventually, the coefficients of thrust ($C_T$) and power ($C_P$) are computed as
\begin{eqnarray}
&& C_T = \frac{T}{\frac{1}{2}\rho\pi (D/2)^2 U_0^2} \label{eq:CT}\\
&& C_P = \frac{P}{\frac{1}{2}\rho\pi (D/2)^2 U_0^3} \label{eq:CP}
\end{eqnarray}

After the computation of hydrodynamic force from ALM at each time step, the force exerted from every Lagrangian point comprising the turbine rotors (as well as the force computed from IBM for nacelles) is transferred back to the fluid grid to correct the Eulerian velocity field.
This interpolation procedure is performed using an isotropic Gaussian projection \citep{Shen2005}, 
\begin{equation}
 f_{L_{ALM}} (x_i) = \frac{1}{\varepsilon^3 \pi^{3/2}} \text{exp} \left( - \frac{r_L^2}{\varepsilon^2}  \right)
\end{equation}
where $r_L$ denotes the radial distance between the marker $L$ and the considered cell face $i$, and $\varepsilon$ is the interpolation stencil set to 3.0$\Delta x_i$.
A Prandtl-type tip-loss correction is adopted to correct the ALM forcing near the blade tip as a function of the number of blades and the tip-speed ratio \citep{Shen2005}.

%%% \newpage\pagebreak

\bibliographystyle{jfm}
% Note the spaces between the initials
\bibliography{0_OuroNishino_v0}

\end{document}